\documentclass[fleqn,usenatbib]{mnras}
\usepackage{graphicx}
\usepackage{multirow} 
\usepackage{dcolumn}
\usepackage{bm}
\usepackage{multicol}
\usepackage{makecell}
\usepackage[utf8]{inputenc}
\DeclareUnicodeCharacter{2212}{-}
\usepackage{longtable}

\usepackage{caption}
\usepackage{lscape}
\usepackage{rotating}

\footnotesize\dimen\footins=0\baselineskip

\title[Passive galaxies at $3<z<7$ with JWST]{Cosmic Stillness: High Quiescent Galaxy Fractions Across Upper Mass Scales in the Early Universe to z = 7 with JWST}

\author[Russell et al.]{Tobias A. Russell,$^{1}$ Neva Dobric,$^{1}$  Nathan J. Adams,$^{1}$ Christopher J. Conselice,$^{1}$, Duncan Austin,$^{1}$ \newauthor Thomas Harvey,$^{1}$  James Trussler,$^{1}$ Leonardo Ferreira,$^{2}$ Lewi Westcott,$^{1}$ Honor Harris,$^{1}$ \newauthor Rogier A. Windhorst,$^{3}$ Dan Coe,$^{9,10,11}$ Seth H. Cohen,$^{4}$ Simon P. Driver,$^{12}$  Brenda Frye,$^{13}$  \newauthor Norman A. Grogin,$^{9}$ Nimish P. Hathi,$^{9}$ Rolf A. Jansen,$^{4}$  Anton M. Koekemoer,$^{9}$ \newauthor Madeline A. Marshall,$^{14,15}$ Rafael {Ortiz~III},$^{4}$ Nor Pirzkal,$^{9}$ Aaron Robotham,$^{12}$ \newauthor Russell E. Ryan,  Jr$^{9}$, Jake Summers,$^{4}$ Jordan C. J. D'Silva,$^{12, 15}$  
Christopher N. A. Willmer,$^{13}$ \newauthor Haojing Yan$^{16}$ 
\\ \\
$^{1}$ Jodrell Bank Centre for Astrophysics, University of Manchester, Oxford Road, Manchester, UK \\
$^{2}$ Department of Physics \& Astronomy, University of Victoria, Finnerty Road, Victoria, British Columbia, V8P 1A1, Canada \\
$^{3}$ School of Earth and Space Exploration, Arizona State University, Tempe, AZ 85287-1404 \\
$^{5}$ University of Louisville, Department of Physics and Astronomy, 102 Natural Science Building, 40292 KY Louisville, USA \\
$^{6}$ Department of Physics, University of the Basque Country UPV/EHU, E-48080 Bilbao, Spain \\
$^{7}$ DIPC, Basque Country UPV/EHU, E-48080 San Sebastian, Spain \\ 
$^{8}$ Ikerbasque, Basque Foundation for Science, E-48011 Bilbao, Spain \\
$^{9}$ Space Telescope Science Institute, 3700 San Martin Drive, Baltimore, MD 21218, USA \\
$^{10}$ Association of Universities for Research in Astronomy (AURA) for the European Space Agency (ESA), STScI, Baltimore,\\ MD 21218, USA \\
$^{11}$ Center for Astrophysical Sciences, Department of Physics and Astronomy, The Johns Hopkins University,\\ 3400 N Charles St. Baltimore, MD 21218, USA \\
$^{12}$ International Centre for Radio Astronomy Research (ICRAR) and the
International Space Centre (ISC),\\ The University of Western Australia, M468, 35 Stirling Highway, \\ Crawley, WA 6009, Australia \\
$^{13}$ Department of Astronomy/Steward Observatory, University of Arizona, 933 N Cherry Ave,
Tucson, AZ, 85721-0009, USA \\
$^{14}$ National Research Council of Canada, Herzberg Astronomy \&
Astrophysics Research Centre, 5071 West Saanich Road,\\ Victoria, BC V9E 2E7,
Canada \\
$^{15}$ ARC Centre of Excellence for All Sky Astrophysics in 3 Dimensions
(ASTRO 3D), Australia \\
$^{16}$ Department of Physics and Astronomy, University of Missouri,
Columbia, MO 65211, USA \\
}

\date{}

\begin{document}
\pubyear{2024}
\maketitle

\begin{abstract}
\noindent
We present a detailed investigation into the abundance and morphology of high redshift quenched galaxies at $3 < z < 7$ using James Webb Space Telescope data in the NEP, CEERS and JADES fields. Within these fields, we identify 90 candidate passive galaxies using specific star formation rates modelled with the BAGPIPES SED fitting code, which is more effective at identifying recently quenched systems than the classical UVJ method. With this sample of galaxies, we find number densities broadly consistent with other works and a rapidly evolving passive fraction of high mass galaxies ($\log_{10}{(M_{\star}/M_{\odot})} > $ 9.5) between $3 < z < 5$. We find that the fraction of galaxies with low star formation rates and mass 9.5 $ < \log_{10}{(M_{\star}/M_{\odot})} < $ 10.5 decreases from $\sim$25\% at $3 < z < 4$ to $\sim$2 \% at $5 < z < 7$. Our passive sample of galaxies is shown to exhibit more compact light profiles compared to star-forming counterparts and some exhibit traces of AGN activity through detections in either the X-ray or radio. At the highest redshifts ($z > 6.5$) passive selections start to include examples of `little red dots’ which complicates any conclusions until their nature is better understood.

\end{abstract}

\section{Introduction}

The role of galaxy quenching, or the truncation of star formation, has always been important for the understanding of galaxy evolution. Our current understanding of galaxies in the local universe allows their separation into two main categories, blue star-forming disk galaxies, and red and “dead” ellipticals often found in galaxy clusters. Some of the first observations of galaxies external to the Milky Way revealed a distinct galaxy morphological categorisation  \citep{Hubble1926}. In a broad sense, a large fraction of galaxies in the nearby universe are elliptical without ongoing star formation, but how and when these galaxies became quiescent is an open question. 

Subsequent investigations revealed a correlation between this phenomenon and the broader-scale surrounding environment of these galaxies \citep{Oemler1974, Dressler1980}. Studies into the morphology of quiescent galaxies such as \citet{Kauffman2003} and \citet{van_der_Wel_2008}, showed that quiescent galaxies are typically more compact and bulge-like than star-forming galaxies. The dynamical process of galaxy transformation into elliptical galaxies is roughly understood through some possible paths derived through simulations \citep{Toomre1972, Mihos1996}, but is likely more complicated than a single route.  While the morphology of these systems can be understood perhaps through dynamical events, we are however still unable to describe why the evolution of ellipticals and spheroids results in a cessation of star formation, transforming a galaxy's colour and activity into red and quenched. 

To prevent exceeding the limits set by galaxy mass functions at lower redshifts \citep[e.g.][]{McLeod2021}, which we would have observed, the quenching of the initial massive galaxies must have occurred at an exceptionally quick pace. However, an alternative explanation suggests a more steady stellar mass assembly at rates typical of galaxies on the main sequence at $z > 4$. This can account for a fraction of the first lower mass quiescent galaxies \citep{Valentino2020}.

Many proposed methods of quenching exist; current theories suggest that sources from inside the galaxies, such as Active Galactic Nuclei (AGN) feedback may drive this. Accreted material onto the supermassive black holes at the centre of galaxies is capable of suppressing the cooling processes and ejecting the gas required for star formation \citep{AGN2_Hopkins}, while emitting substantial energy across the electromagnetic spectrum, including X-rays and radio waves.  This energy can furthermore heat any other existing gas, making it difficult for this gas to cool and form new stars.  This can act to effectively transform blue star-forming spiral galaxies into red, quenched galaxies (eg. \citet{Chen_2020}).  The study by \citet{Bluck_2023}, showed that the stellar gravitational potential is the most important parameter of AGN that affects quenching. \citet{AGN4_Dubois} confirmed through simulations of haloes that in the absence of AGN feedback, large amounts of stars accumulate in the central galaxies to form overly massive, compact, rotation-dominated galaxies - which are not observed - resulting in a consensus that supermassive black holes play an important role in suppressing star formation in these galaxies.

There is also evidence for quenching driven by factors from the surrounding environment, particularly in the context of galaxy clusters, such as ram pressure stripping. Ram pressure stripping describes the process of the Intra-Cluster Medium (ICM), which permeates the space between galaxies in dense environments, exerting kinetic pressure on the gas in galactic reservoirs, which effectively strips them of the cold gas required for new star formation \citep{Maier2019}. This effect is stronger in galaxy clusters, as the denser environment produces a more significant interaction with the ICM. Work done by \citet{Zinger2018} shows that the ICM effects quenching primarily in galaxies located at the cores of a galaxy cluster, while satellite galaxies experience delayed quenching compared to the central galaxies.

Galactic mergers can also serve as a quenching mechanism, even though they are often associated with enhanced star formation due to gravitational interactions. If two spiral galaxies collide at the right orientation and velocity, they merge in a manner that expels a significant portion of dust and gas and primarily adds mass to the galaxy centre \citep{lambas_2012}, driving a strong circumnuclear starburst that simultaneously feeds an AGN. This results in rapid heating and removal of the cold gas to leave a quenched elliptical galaxy with a large black hole, as found by for example in \citet{Alexander_2012}.
 
Secular processes, characterised by slow and non-violent structural changes also contribute to quenching. Stellar feedback, in which stars release energy and matter into the interstellar medium, can influence the redistribution of gas, altering the conditions of ongoing star formation \citep{Silk_2012}. A prominent example of this is supernova feedback, which is caused by shocks and heating from supernova explosions impacting the gas reserves in galaxies  \citep{Gelli_2023}. However, this is probably only relevant for low-mass galaxies, as found by \citet{Chan2018}. This is because low-mass galaxies have smaller potentials, so lower energy processes like supernovae have a greater impact. In the case of more massive galaxies other types of feedback, such as AGN and central supermassive black holes, are needed to quench galaxies \citep{Su2019}.

While quenching can be environmental or mass-driven, the prevailing theory is that mass quenching is the dominating influence on massive galaxies \citep[e.g.,][]{Bluck_2023}, particularly at high redshift \citep{Contini2020}. As quenching has been discovered to be a key phase in the life of massive galaxies \citep{Balogh2004, Faber2007}, there is a strong incentive to identify as many early massive quiescent galaxies as possible and use these to study what drives this process, and when it occurs.

Observationally, it was once thought that more evolved galaxies in the distant universe would be extremely rare if not non-existent \citep[e.g.,][]{Bahcall1990}.  From the outset of deep HST imaging, it was thought that many observed distant galaxies were peculiar in morphology, likely driven by the merger process \citep{Driver1995, Conselice2003}.   However, we now know from early JWST observations that galaxies are much more evolved than expected, finding early type and disk galaxy morphologies in a higher abundance at higher redshifts, compared with HST observations \cite[e.g.,][]{Ferreira2022,ferreira2023jwst}.   In fact, it has been shown from these early studies that galaxy formation is quite advanced in terms of the formation of galaxy structure. 

The direct search for quiescent galaxies in the early Universe commenced with a study done by \citet{Fontana_2009} where quiescent galaxies at $z > 2.5$ were discovered based on their specific star formation rates from spectral energy distribution fitting.  Recently, the search intensified with the Spitzer Space Telescope \citep{Houck_2004, Carnall2019, Spitzer2011}, and more recently by the James Webb Space Telescope (JWST). The JWST has provided unparalleled resolution and sensitivity, enabling deep infrared imaging beyond $\lambda > 2 \mu$m. This has opened up the opportunity to look back at more distant epochs than ever before. JWST also provides precise angular resolution to probe finer morphological details of galaxies \citep{Katherine}, and an extensive range of spectroscopic capabilities \citep{Carnall2022}. These promise for the unprecedented detailed analysis of individual galaxy properties, such as the star-formation history, stellar content metallicity, and sizes. 

JWST provides the opportunity to identify and study large numbers of quiescent galaxies through the combination of the modelling of galaxy Spectral Energy Distributions (SEDs) and its spectroscopic capabilities. This has been performed at redshifts $z > 3$ with number densities estimated by \citet{Carnall2019} and \citet{Valentino2023}. With its spectroscopic capabilities extending beyond the near-infrared atmospheric window, JWST is able to reliably confirm galaxy quiescence beyond $z > 4.5$ \citep{carnall2023, degraaff2024}, and it has even identified low star formation rates up to $z \sim 7$ \citep{looser2024recently, trussler2024like, weibel2024}. There is also strong evidence to suggest quiescent galaxies can be found via spectral energy distribution at redshifts $z > 5$. Current research at $3 < z < 4$ \citep{carnall2023, Valentino2023, Girelli2019} reveals a higher number density of quiescent galaxies than indicated by current simulations and models \citep{Cecchi_2019}. This suggests that some of the physics is still missing from our current models describing the rapid quenching events in the early Universe. Moreover, theoretical predictions like the SHARK model \citep{SHARK} and $Illustris$-TNG simulation \citep{Illustris} find fewer massive quenched galaxies at $3.5 < z < 4.5$ than current observations show. Both creating enough massive galaxies in the early Universe and quenching them such that they are passive without star formation on a short timescale remains a challenge for current galaxy formation models.

However, the number of these distant sources remains small and more fields need to be analysed to find these quenched systems at a range of masses. As such, in this paper, we identify quiescent candidates at stellar masses $M_{*} > 10^{8.5}M_\odot$ and at redshifts $z > 3$  using a compilation of NIRCam imaging surveys (with the PEARLS, CEERS, and JADES fields). In this work, we measure how the number density of these systems evolves with time and how their morphological properties (S\'{e}rsic index, half-light radius) differ relative to the star-forming population. Employing the \texttt{BAGPIPES} code \citep{Carnall2018, Carnall2019}, used for spectral fitting, we extract specific star formation rates, mass, and age parameters to differentiate between quiescent and star-forming galaxies. 

The structure of this work is as follows. In section \ref{data} we cover the field data used and calibrations made to this prior to this analysis. Section \ref{method} covers the methods used in identifying passive candidates and calculating number density values. In section \ref{results} we present our results. Section \ref{discussion} discuss their implications. Finally, Section \ref{conc} presents our conclusions and summarises the paper. As a note, we use the terms quenched, passive and quiescent galaxy as interchangeable terms within this paper.  For cosmological calculations, we adopt $\Omega_{M}$ = 0.3, $\Omega_{\Lambda}$ = 0.7 and H$_{0}$= 70 km s$^{-1}$ Mpc$^{-1}$. All magnitudes given in this work follow the AB magnitude system \citep{Oke1974, Oke1983}.

\section{Data} \label{data}

We make use of data taken from imaging by the Near Infrared Camera, (NIRCam, \citet{nircam_overview, Rieke2008, Rieke_2015, Rieke_2023}).   This is our primary source of data, although we use other data, such as spectroscopy and multiwavelength detections, to study the objects which we identify as potential quiescent galaxies. 

\subsection{NIRCam Near-Infrared Data} 

In this work, we use data from several different fields, including the Cosmic Evolution Early Release Science (CEERS, PID: 1345, PI: S. Finkelstein, \citet{Bagley_2023}) survey, the North Ecliptic Pole Time Domain Field (NEP-TDF, or NEP) located within the Prime Extragalactic Areas for Reionization Science (PEARLS, PID: 1176 \& 2738, PI: R. Windhorst \& H.Hammel, \citet{NEP_mag}), and the JWST Advanced Deep Extragalactic Survey Data Release 1 (JADES DR1, PID: 1180, PI: D. Eisenstein, \citet{eisenstein2023overview}). These fields contain images consisting of  10, 8 and 6 NIRCam pointings respectively. The CEERS dataset employs specific NIRCam photometric filters, including F115W, F150W, F200W, F277W, F356W, F410M, and F444W. Additionally, we use HST data from the F606W and F814W bands of the Cosmic Assembly Near-infrared Deep Extragalactic Legacy Survey (CANDELS) field HDR1 version, as documented by \citet{Koekemoer_2011} and \citet{Bagley_2023}. The NEP data incorporates NIRCam filters F090W, F115W, F150W, F200W, F277W, F356W, F410M, and F444W, as well as yet unreleased HST data covering F606W (Jansen et al. (In Prep)). The JADES data uses the F090W, F115W, F150W, F200W, F277W, F335W, F356W, F410M, and F444W NIRCam filters and we include HST data in the F606W band from the Hubble Legacy Fields Project \citep{Whitaker_2019}.  A detailed description of these fields and our particular reduction methods and analysis can be found in \citet{Adams2023, Harvey2024, Conselice2024}.

The CEERS field consists of 10 NIRCam pointings covering the Extended Groth Strip and covers a total area of 61.42 arcmin$^2$ after masking. NEP covers 8 NIRCam pointings in the JWST continuous viewing zone near the North Ecliptic Pole and covers a total area of 57.32 arcmin$^2$ after masking the original region \citep{Jansen2018}. JADES covers 6 overlapping NIRCam pointings around the Hubble Ultra Deep Field in GOODS-South and covers a total area of 22.98 arcmin$^2$ after masking. Our masking covers a small buffer around image edges, the diffraction spikes of bright stars and any residual artefacts identified by eye in the imaging. The depths of these surveys in their respective filter sets are discussed in detail in \citep[]{Conselice2024, Harvey2024}. 


\subsection{Calibration and Treatment in Prior Work} \label{Nathandata}

This work makes use of the EPOCHS series of data reductions, ensuring each survey has been reduced in the same way \citep{Adams2023}. To summarise the process,  uncalibrated lower-level JWST data was processed using version 1.8.2 of the official JWST reduction pipeline and Calibration Reference Data System v1084. Wisps and other artefacts in the data are subtracted in this process. The identification and extraction of sources were performed by \texttt{SExtractor} \citep{BertinandArnouts1996}, providing the fluxes and corresponding magnitudes of different measured aperture sizes centred on sources for each waveband.

Source identification is conducted on a weighted stack of the red broad bands. Photometry is calculated using  0.32 arcsec diameter circular apertures. Aperture-corrected values are included based on corrections derived from simulated \texttt{WebbPSF} Point Spread Functions (PSFs) for each band \citep{Webbpsf2014}. This corrects the measured flux by a factor to account for using a circular aperture, potentially smaller than the full size of a source.

Many of the sources examined in this work, while small, are resolved and not point sources. Aperture photometry is used to measure galaxy colours in order to conduct SED fitting with a high S/N ratio. When examining physical properties, we correct the normalisation of our galaxy SED's using SExtractor's MAG\_AUTO measurement in F444W in order to capture emission missed by our small apertures. Finally, the local depths across each image in each field were found by placing circular apertures in empty regions of the image. Here we take `empty' to mean that no pre-existing sources were found within 1 arcsec of the central aperture coordinate by \texttt{SExtractor}.

\subsection{Ancillary Data}

The fields we use have been well covered by studies 
featuring multi-wavelength and spectroscopic data which we conduct comparisons to at various points. These include the following:

\subsubsection{Mid-Infrared Data}

We make use of Mid-Infrared Imager (MIRI) data partially covering the CEERS field; coverage of the NEP and JADES fields is unavailable. The Mid-Infrared (MIR) data covers the wavelength range 5 – 20 $\mu$m \citep{yang2023ceers}, which is a much redder range than the 0.6 - 5 $\mu$m covered by NIRcam \citep{Rieke_2023}. The full data reduction process is described in \citet{yang2023ceers}, which should be consulted for details.

\subsubsection{Spectroscopic Data}

We also use spectroscopic data in our analysis. We make use of 3D-HST Treasury Survey data which covers the CEERS field, including 4 out of 5 of the deep fields observed by CANDELS This data was reduced using a similar procedure to the CANDELS team \citep{momcheva20163d}. Alongside this data, there is spectra from the MOSFIRE Deep Evolution Field (MOSDEF) Survey performed by \citet{kriek2015mosfire} which we also us. We also incorporated data from the JADES data release 1 spectra taken by NIRSpec as part of the JWST Deep Extragalactic Survey done by \citet{bunker2023jades}. We also compare our CEERS field photometric redshifts to data from the v3 release from the Dawn JWST Archive (DJA) \citep{heintz_2024}, including spectra from the recent RUBIES programme by \citet{DeGraaf2024}. The individual observing programmes belonging to this, whose data we use, are discussed in further detail in Section \ref{multiwave}.

\subsubsection{X-ray Data}

In this work, we compare our findings to data from X-ray catalogues covering the same fields. We use the X-ray Chandra Deep Field South Survey to supplement the JADES data \citep{luo2016chandra}. This survey has a flux limit of $\sim$ 9 $\times$ 10$^{-16}$ ergs$^{-1}$ cm$^{-2}$ in the 0.5 - 10 keV bands. 
For the NEP field we used NuSTAR and the XMM-Newton extragalactic survey data from \cite{zhao2024pearls}, which have sensitivities of $\sim$ 3 $\times$ 10$^{-14}$ ergs$^{-1}$ cm$^{-2}$ in the 3 - 24 keV bands and $\sim$ 6 $\times$ 10$^{-15}$ ergs$^{-1}$ cm$^{-2}$ in the 0.5 - 10 keV bands respectively. We used deep Chandra imaging from the AEGIS-X Deep survey data by \cite{nandra2015aegis} to provide X-ray coverage for the CEERS field, which has a sensitivity of $\sim$ 2 $\times$ 10$^{-16}$ ergs$^{-1}$ cm$^{-2}$ across the 0.5 - 7 keV bands.

\subsubsection{Radio Data}

We also make use of Radio data catalogues covering the CEERS, NEP and JADES fields. We make use of 1.4 Ghz AEGIS20 radio survey data taken using the Very Large Array (VLA), which covers the CEERS field. Full information on observations and data treatment can be found in \citet{ivison2007aegis20}. The NEP field data was extended by imaging performed by \citet{hyun2023jcmt} using the James Clerk Maxwell SCUBA-2 telescope which operated at 850 $\mu$m. As this data was initially presented alongside supplementary imaging performed by the VLA at 3 Ghz, we also include this.  The JADES field radio imaging taken by the VLA at 1.4 GHz \citet{miller2013very}.

\section{Methods}\label{method}

In this section, we describe the process of isolating our sample of quiescent galaxies, fractions of quiescent galaxies, as well as the method of calculating comoving number densities. Following this, we analyse the morphology of our quiescent sample. The initial sample across all 3 fields contains 209,009 identified sources \citep{Adams2023}.

We start by analysing the initial sample using \texttt{EAZY}, which gives us the photometric redshifts and magnitudes of galaxies, but it does not give physical properties of the galaxies. After filtering galaxies based on their redshifts and magnitudes, we analyse the remaining sample using \texttt{Le PHARE}. \texttt{Le PHARE} gives us the physical properties of galaxies, like stellar mass, SFR, sSFR, metallicity and others. Based on sSFR, we determine possible quiescent sample. We analyse the sample further using \texttt{BAGPIPES} which gives us values of the same physical properties and their uncertainties with higher precision than \texttt{Le PHARE}, 
but its Bayesian approach enables a better quantification of the potential uncertainties on these physical properties; however, this is a more time-consuming analysis.

\subsection{Initial Cuts to \texttt{EAZY} data}\label{EAZYdatamethod}

We use \texttt{EAZY} to measure initial photometric redshifts for our sample \citep{EAZY}. We run \texttt{EAZY} using the template sets presented in \citet{Larson_2023} with dust extinction up to 3.5 magnitudes \citet{Calzetti} and the \citet{Madau1995} treatment for absorption by neutral hydrogen. We make use of the aperture-corrected fluxes and corresponding magnitude values from each waveband. 

We begin our cuts with a 5 $\sigma$ cut in the reddest photometric band, F444W. This removes faint objects, as noisy objects in the reddest band will have too poor of a signal for analysis in other bluer bands. We follow this with a cut in the best fitting photometric redshift of $z > 2.5$. This redshift cut is used as our later analysis with BAGPIPES allows redshift to remain free, resulting in some scatter around $z = 3$ which we use as a lower redshift cut later.

For the few objects we identify at very high redshifts ($z > 6.5$) we limit the sample to those selected in \citet{Conselice2024} which employs additional criteria (e.g. on Lyman break strengths and morphology) to increase selection reliability. In addition, we fit templates of brown dwarfs to these higher redshfit galaxies using the Sonora Bobcat models \citep{Marley2021,Harvey2024} and remove those better fit as such. These cuts reduce our sample size from 209,009 to 28,385 objects. 

\subsection{Le PHARE}\label{lepharemethod}

The \texttt{Le PHARE} code computes  photometric redshifts and other galaxy parameters, such as SFR, sSFR and mass, as well as to perform Spectral Energy Distribution (SED) fitting \citep{arnouts-1999,2006Ilbert, Arnouts1999}.   We use \texttt{Le PHARE} to measure the stellar masses and star formation rates of our galaxies as a first look. 

Within SEDs the Balmer break is a characteristic feature of massive quiescent galaxies in this redshift region. The Balmer break occurs at the Balmer limit, with a rest wavelength of $\lambda$ = 3645 \AA  \citep{Wilkins2023}. This feature is present due to hot stars, such as A-class stars \citep{Bessell2007}. The magnitude of the Balmer break is proportional to the abundance of A-type stars and inversely proportional to the amount of O- and B-type stars within a galaxy. As the latter stars possess comparatively shorter lifespans \citep{Weidner2010}, there will be fewer of these sources in an environment with no recent star formation activity, hence the Balmer break is a pronounced feature in the spectra of massive quiescent galaxies.

Following our redshift cuts, the remaining sample was run through \texttt{Le PHARE} using the \citet{BC03} template models for computing the spectral evolution of stellar populations. These are theoretical spectra covering fitting parameter ranges which vary according to the different kinds of observed galaxies, and act as a guideline for the final parameters to be fitted to. The redshift for this run was fixed to the best template model fit by \texttt{EAZY}.  Thus we do not use \texttt{Le PHARE} for the photometric redshift measures, but only as an initial measure of the physical properties of the galaxies in our sample. 

To isolate quiescent galaxies from our sample, we chose the identification method of limiting sSFR, following a similar analysis done in \citet{carnall2023}. We use this method of identifying quiescence as the JWST is not able to probe the rest-frame J band at high redshifts, as well as due to limitations with the classic UVJ selection method for young quiescent galaxies in the early universe, see Section \ref{uvj} for further discussion. We make use of a time-dependent cut, as has been widely used in literature (for example \citep{Gallazzi_2014, 2016ApJ...832...79P},

\begin{equation}
    \label{eq:sSFR}
    sSFR < \frac{0.2}{t_{obs}}
\end{equation}
where $t_{obs}$ represents the age of the Universe at a given redshift. $t_{obs}$ can be found following \citet{1993_peebles}, using 
\begin{equation}
    t_{obs} = \frac{2}{3 H_{0} \sqrt{\Omega_\Lambda}} \sinh \left( \left( \frac{1}{1+z}\right)^{3/2}  \sqrt{ \frac{\Omega_\Lambda}{\Omega_m}} \right) ^{-1}
\end{equation}
where $ H_{0} $ is the Hubble constant, $\Omega_\Lambda$ is the fraction of dark energy in the total energy density of the universe, $\Omega_m$ is the fraction of matter in the total energy density of the universe and $ z $ is the redshift. 

To account for the cases when objects have \texttt{Le PHARE} fit sSFR values which scatter around selection limits, we implement an initial cut which is less strict
\begin{equation}
    \label{eq:sSFR_loose}
    sSFR < \frac{0.3}{t_{obs}},
\end{equation}
before our sample moves to \texttt{BAGPIPES} fitting. When not utilising this more lax cut, we miss out on around 5\% of the objects in our final sample, most commonly low-mass objects. 

There is an imprecisely known level of systematic error introduced to the flux values through various sources. These include unknown errors in the images and method of building the catalogues and systematic offsets in JWST measurements. We also consider that the templates used in the \texttt{Le PHARE} fitting process are imperfect, so allowing for larger error values allows for a better minimisation of the $\chi^2$ fit value. The best estimates as of writing place JWST absolute flux calibration for NIRCam at better than 5\% \citep{ma2024jwsts}. To minimise the chance of underestimating errors we set a minimum error of 10\% on flux values. 

These initial, and relaxed, sSFR cuts using \texttt{Le PHARE} reduce our sample to 312 objects which we proceed to model using \texttt{BAGPIPES}.

\subsection{BAGPIPES}\label{BAGIPESMETHOD}

In order to fit spectra to our data, we make use of Bayesian Analysis of Galaxies for Physical Inference and Parameter Estimation (\texttt{BAGPIPES}) \citep{Carnall2017}. This is a \texttt{Python} code which is able to rapidly produce detailed galaxy spectra, and then fits these to variable combinations of photometric and spectroscopic data. It is also able to provide information on various galaxy properties, such as stellar mass, SFR, SFH, and redshift. \texttt{BAGPIPES} does this by utilising the MultiNest Nested Sampling algorithm created by \citet{Skilling2004}, which uses principles from Bayesian statistics. Additionally, \texttt{BAGPIPES} algorithm uses a stellar evolutionary track developed in \citet{Bressan2012} and \citet{Marigo2013}, as well as models of stellar populations from work by \citet{BC03} and their updated version by \citet{Chevallard2016}. This can be utilised to fit galaxy spectra to \texttt{EAZY} catalogue data, employing the flux and its corresponding errors from all available wavebands. The MultiNest algorithm is able to provide a more detailed picture of model parameter errors than the best-fit results \texttt{Le PHARE} generates at the cost of a longer running time. \texttt{BAGPIPES} uses the prior probability to find the mass formed, star formation rate, specific star formation rate, galaxy age, dust attenuation, and the photometric redshift.

At this stage, we also apply additional 5$\sigma$ cuts to the F200W and F277W bands, as these are the first bands located blueward of the Balmer break for galaxies at $z\approx 3-4$ and $z > 4$ respectively. These ensure that strong Balmer breaks are present in objects passing our selection cuts.

Our model choices largely follow those in \citet{carnall2023}. The total stellar mass was given a logarithmic prior, which ranges from $10^{5}$ to $10^{12}$ Solar masses. Nebular line and continuum emission are included by varying the ionisation parameter $U$ from $10^{-2}$ to $10^{-3}$, following a modification to work done by \citet{ionisation}. Stellar and gas-phase metallicities have logarithmic priors and can range from 0.2 to 2.5. The age of the galaxy has a uniform prior. Dust attenuation is included using the model by \citet{salim2018dust} which has a variable slope described by the Calzetti model \citep{Calzetti}, where we set a uniform prior. Redshift is fitted photometrically by \texttt{BAGPIPES}, with the uniform prior and it can vary from 0 to 20. 
For the purpose of this analysis, we exclude all galaxies with redshifts of $z < 3$ fitted by \texttt{BAGPIPES} as these are lower than the regime we are interested in this paper.

\texttt{EAZY} fitted redshift values differ from those fitted by \texttt{BAGPIPES} on average by a magnitude of 0.616 $\pm$ 0.529 in our sample, with \texttt{EAZY} fitting higher values on average. In later comparisons to spectroscopic data, we find the \texttt{BAGPIPES} photometric redshifts to be more robust.

\subsection{sSFR: Quiescent and Robust Samples}

Following previous work on this topic, including \citet{carnall2023}, we aim to concretely quantify whether a galaxy is quiescent or not in a more sophisticated way for both redshift samples. To do this, the distributions of the sSFR PDF data output by \texttt{BAGPIPES}, were used. We set a limiting value to selected quiescent systems based on the sSFR cut given in Equation \ref{eq:sSFR}; at least 50\% of the area of each galaxiy's sSFR PDF must be below this limit for the system to be considered passive. This method is used instead of the raw values given by \texttt{BAGPIPES}, as using the individual values may miss a significant secondary star-forming solution in the sSFR PDF.

 
We also create a second more strict, robust subset of data of our passive candidates. This required 97.5\%, instead of 50\%, of the PDF area to lie below the limit established by the sSFR cut value in Equation \ref{eq:sSFR}. This allowed the exclusion of objects with any significant star-forming solutions with a higher level of confidence.

\begin{figure*}
    \centering
\includegraphics[width=0.8\textwidth]{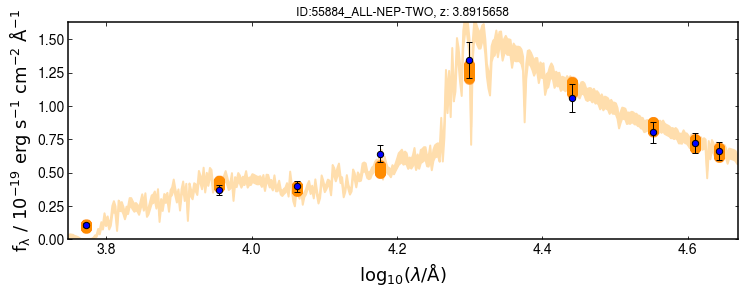}
    \caption{Example of fitted spectral energy distribution to the galaxy in NEP field with the unique ID 55884, whose redshift is fit by \texttt{Le PHARE}. The black points represent measurements from the JWST after calibration, and the orange line represents the fitted SED. Also, the lighter orange line represents the whole fit, while the darker orange marker represents where the measured band is being fitted. The Balmer break is clearly visible, and the fitted SED after the Balmer break is within our measured points, hence this is an example of an object passing the visual inspection.}
    \label{fig:Balmer}
\end{figure*}
Before calculating our final number densities, the sample was cleaned by visually inspecting the fitting of the photometric bands, as well as the \texttt{BAGPIPES} SEDs. This brought our full passive sample size from 113 after cuts based on sSFR to 90 after visual inspection. Figure \ref{fig:Balmer} displays an example of a SED that made our final selection cut. Incorporating the visual inspection of SEDs into our previous cuts allowed us to get our final full passive and robust passive samples. 

\subsection{Assessing our Selection Criteria}
\label{compl}

In order to determine if our selection procedure for passive galaxies is reliable, we utilise mock catalogs of galaxies using the JAGUAR semi-analytic simulation \citep{Williams18}. The goal is to assess which galaxies truly meet our selection criteria and determine if we would select them if we conduct mock observations and run them through our selection pipeline. Our use of these models is also explored in detail in \citet{Conselice2024} and \citet{Adams2023}.

We use three realisations of these JAGUAR simulations, providing approximately 300 square arcminutes of simulated area. Three catalogs are then generated by scattering the photometry using the mean depths of the three fields from this study (CEERS, NEP, JADES). The resultant catalogs are run through our SED modelling and selection pipeline. We find that the completeness rate, that is the number of objects we detect correctly as passive systems, at $z>3$ is relatively high at 79\%, with 14\% of sources lost as photo-z outliers and 7\% of objects lost due to failing our criteria of requiring at least 50\% of the sSFR PDF below a certain limit. For the objects selected, our masses are systematically smaller by 0.13~dex, with a scatter of 0.11~dex. Photo-z outliers are driven by the red colours of these passive galaxies, with a $\sim3\%$ outlier rates for those with $m_{\rm F277W}<25$ and increasing towards fainter magnitudes, with zero successful identifications at $m_{\rm F277W}>27.5$ in JADES and $m_{\rm F277W}>26.5$ in both CEERS and NEP. This is due to their very faint bluer colours at $\sim1\mu$m resulting in poorer SED fits. We find that completeness dramatically falls off at stellar masses of around $10^9 M_\odot$ and below. Thus, in conclusion from this test, we are recovering most of the bright, massive and quiescent systems at high redshifts, although we are less successful for the fainter systems. 

\subsection{Quiescent Galaxy Sample}\label{dens_calc}
\footnotetext[1]{https://www.ph.unimelb.edu.au/~mtrenti/cvc/CosmicVariance.html}

We investigate quenched galaxy frequency by calculating the comoving number densities for quenched systems in each of our fields. To calculate and meaningfully compare number densities with previous literature, the samples are divided into redshift and mass bins. We used the redshift ranges $3 \leq z < 4$, $4 \leq z < 5$, and $5 \leq z < 7$ to investigate the evolution in the number densities of these systems. This allows us to study the evolution of quenched galaxies and their number density with redshift. We separate results into three samples based on their stellar masses, with a high and a low stellar mass cut used. The low mass galaxies are those with values $\log_{10}{(M_{\star}/M_{\odot})}<9.5$, medium mass galaxies have masses $9.5\leq\log_{10}{(M_{\star}/M_{\odot})}<10.6$, and high mass galaxies are those where $10.6\leq\log_{10}{(M_{\star}/M_{\odot})}$.

We identify a total of 14 high mass ($10.6\leq\log_{10}{(M_{\star}/M_{\odot})}$) galaxies in the full sample, and 12 high mass galaxies in the robust sample. We find a total of 52 medium mass ($9.5 \leq\log_{10}{(M_{\star}/M_{\odot})}< 10.6$) galaxies in the full sample, and 25 medium mass galaxies in the robust sample. Finally, we identify a total of 21 low mass ($\log_{10}{(M_{\star}/M_{\odot})}<9.5$) galaxies in the full sample, and 4 low mass galaxies in the robust sample. These results are shown in Table \ref{table:RESULTS}

The uncertainties on the number density values were acquired through computing the Poisson noise on the number of galaxies identified with an approximation,
\begin{equation}
\label{eq:Pois}
    \sigma_{Poisson} = \frac{\sqrt{N}}{N}
\end{equation}
where $N$ refers to the number of massive quiescent galaxies found in the field. We refer to the full equation in \citet{Gehrels1986} for use on small samples containing less than 5 galaxies, where this approximation breaks down. We also consider the error introduced through the effects of cosmic variance and calculate this value for each field using the Cosmic Variance Calculator\footnotemark[1], which follows the method originally laid out in \citet{Trenti_2008}. This provides 1$\sigma$ fractional uncertainties on the number counts, supplying the total uncertainty, as well as the separate Poisson noise and cosmic variance values. We assume completeness and halo-filling factors of 1 and thus choose the intrinsic number of objects to be the total number of galaxies identified for that redshift bin. For the full method of cosmic variance calculation,  see \citet{Trenti_2008}. Following \citet{Valentino_2023}, we use the method for combining cosmic variance originally laid out in \citet{Moster_2011}. The total cosmic variance values, $\sigma_{CV,total}$, for the combined field were found through the use of
\begin{equation}
    \sigma_{CV,total} = \sqrt{\frac{1}{\Sigma_{fields} \sigma_{CV,fields}^{-2}}} 
\end{equation}
where it is assumed that all fields are independent of one another.

\begin{table*}
    \renewcommand{\arraystretch}{1.5}
    \caption{The full and robust samples of comoving number densities derived from quiescent galaxies in redshift bins spanning 3 $\leq$ z $<$ 4, 4 $\leq$ z $<$ 5, and 5 $\leq$ z $<$ 7. We display estimates from our higher mass cuts, $M \leq 10^{10.6} M_{\odot}$, and middle mass cuts, $10^{9.5} M_{\odot} \leq M < 10^{10.6} M_{\odot}$, and lower mass cuts $M \leq 10^{9.5} M_{\odot}$.}
    \centering
    \label{table:RESULTS}
    \begin{tabular}{p{0.18\linewidth}p{0.07\linewidth}p{0.1\linewidth}p{0.15\linewidth}p{0.1\linewidth}p{0.15\linewidth}}
        
        \hline
          &  & & \textbf{Full Sample}& &  \textbf{Robust Sample} \\
        \hline
        \textbf{Mass Cut} &\textbf{Redshift Range} & \textbf{Number of Galaxies} & \textbf{Number Density (Mpc$^{-3} \times 10^{-5}$)}& \textbf{Number of Galaxies} & \textbf{Number~Density (Mpc$^{-3} \times 10^{-5}$)} \\
        \hline
        \hline
        $M \geq 10^{10.6} M_{\odot}$ & 3 $\leq$ z $<$ 4 & 10 & 2.19$^{+0.81}_{-0.81}$ & 9 &  1.97$^{+0.76}_{-0.76}$\\
        &4 $\leq$ z $<$ 5 & 1 & 0.24$^{+0.56}_{-0.22}$ & 1 &  0.24$^{+0.56}_{-0.22}$\\
        &5 $\leq$ z $<$ 7 & 3 & 0.43$^{+0.45}_{-0.29}$ & 2 &  0.29$^{+0.38}_{-0.20}$\\
        \hline
        $10^{9.5} M_{\odot} \leq M < 10^{10.6} M_{\odot}$ & 3 $\leq$ z $<$ 4 & 28 & 6.14$^{+1.53}_{-1.53}$ & 14 &  3.07$^{+0.99}_{-0.99}$\\
        & 4 $\leq$ z $<$ 5 & 19 & 4.60$^{+1.37}_{-1.37}$ & 8 &  1.94$^{+0.82}_{-0.82}$\\
        & 5 $\leq$ z $<$ 7 & 5 & 0.71$^{+0.35}_{-0.35}$ & 3 &  0.43$^{+0.42}_{-0.25}$\\
        \hline
        $M \leq 10^{9.5} M_{\odot}$ & 3 $\leq$ z $<$ 4 & 13 & 2.85$^{+0.95}_{-0.95}$ & 3 &  0.66$^{+0.66}_{-0.40}$\\
        & 4 $\leq$ z $<$ 5 & 5 & 1.21$^{+0.60}_{-0.60}$ & 1 &  0.24$^{+0.56}_{-0.22}$\\
        & 5 $\leq$ z $<$ 7 & 3 & 0.43$^{+0.42}_{-0.25}$ & 0 &  0.00$^{+0.26}_{-0.00}$\\
        \hline
    \end{tabular}
\end{table*}

\subsection{GALFIT analysis}

We conduct a more detailed analysis of the morphology of our sample, by making use of \texttt{GALFIT} version 3.0.5 developed by \citet{Galfit} and \citet{Peng_2010}, which is a two-dimensional fitting algorithm used to obtain structural components as they appear in two-dimensional galaxy images and fit the light profiles to the images. It is used to fit parametric functions such as Sérsic profiles to galaxies, which we do for our sample.  \texttt{GALFIT}  also allows for the simultaneous fitting of an arbitrary number of components and a combination of the functional forms mentioned previously.  

\texttt{GALFIT} uses a least-squares fitting algorithm to capture the surface brightness profiles of galaxies. This software requires a "data" image to measure the surface brightness and a "sigma" image, giving the relative error at each pixel in the image; from these, the model image is calculated. 

We ran \texttt{GALFIT} individually on the F444W band images of all galaxies passing visual inspection from our final sample. We analysed cutouts of quiescent candidates determined by the \texttt{BAGPIPES} run in the F444W band, as that is the reddest band allowing us to trace as much as possible the underlying stellar mass with the highest S/N. This band, at the high redshifts, also corresponds to observed rest-frame optical light, thus this also enables a comparison to lower redshift optical morphologies. 

To carry out these measurements with \texttt{GALFIT} we create 101 by 101 pixel image cutouts of our sources, as this is of sufficient size to fully contain each object centred in the middle of the image. By observation, we discovered that many images contained multiple objects. This posed an issue as the \texttt{GALFIT} fitting process is highly sensitive and often encounters errors when multiple objects are located in the same image, as it attempts to fit all objects simultaneously.
To combat this issue we masked out other galaxies in these images using a segmentation map produced by \texttt{SExtractor} in \citet{Adams2023}. As we are using the F444W band, we made use of the corresponding PSFs in the fitting process. In the input configuration, as we required a "sigma" image for \texttt{GALFIT}, we used a 1$\sigma$ level noise image, the ERR extension, provided by the JWST pipeline. The software also requires an input estimate for the x and y image coordinates, magnitude, half-light radius, axis ratio, position angle and S\'{e}rsic index. As done in \citet{Katherine} and \citet{kartaltepe2023ceers}, we use the results of the \texttt{SExtractor} catalogue for our initial parameter estimates. We used the \texttt{FLUX$\_$RADIUS$\_$F444W} column as the initial value for fitting the effective radius, as this is the half-light radius given by \texttt{SExtractor}. We used the F444W magnitude values produced by \texttt{SExtractor} for the total magnitude values. 

We fit a Sérsic profile to our systems using \texttt{GALFIT}, which has the form:
\begin{equation}
    \label{eq:Sersic}
    I(R) = I_0 \exp \Biggl\{ -k \left[ \left( \frac{R}{R_0} \right)  ^{1/n}-1 \right] \Biggr\}
\end{equation}
where $I_0$ is the intensity at the centre and $k$ is a dependent variable coupled to $n$, the S\'{e}rsic index.  The S\'{e}rsic index describes how rapidly the light intensity changes with the radius. A large value indicates a steep inner profile and an extended outer region, and a small value indicates a shallow inner profile with a steeper cutoff at a large distance. Galaxies with low values are often disc-dominant or ongoing mergers. A greater index indicates a more centrally concentrated light profile, such as for elliptical galaxies.

\texttt{GALFIT} outputs several parameters such as the galaxy magnitude, half-light radius, Sérsic index, and reduced $\chi ^ {2}$ of the fit. Alongside these it produces three images, the galaxy model, the background emission and the residual image. 

A residual image is obtained by subtracting the model and the background image from the real image. We used these properties to further analyse the structures of our galaxies. To assess the quality of the model galaxy obtained by \texttt{GALFIT}, we calculate the so-called residual flux fraction. The residual flux fraction (RFF) represents the portion of the signal within the residual image that remains unexplained by background fluctuations \citep{Hoyos_2011}. Therefore, a lower RFF indicates a more accurate fitting. RFF is calculated following \citet{RFF}:
\begin{equation}
\label{eq:RFF}
\mathrm{RFF}=\frac{\sum_{(j, k) \in A}\left|I_{j, k}-I_{j, k}^{\mathrm{GALFIT}}\right|-0.8 \sum_{(j, k) \in A} \sigma_{\mathrm{B} j, k}}{F L U X_{-} A U T O},
\end{equation} 
where $\sum_{(j, k) \in A}\left|I_{j, k}-I_{j, k}^{\mathrm{GALFIT}}\right|$ is the sum of all emission within the residual image output from \texttt{GALFIT}, $\sigma_{\mathrm{B} j, k}$ is the background emission and FLUX$\_$AUTO is the sum of the emission. We measure the emission for these quantities using apertures defined as having a 0.5-arcsecond diameter, which has been chosen to accurately encapsulate the size of most galaxies in our sample.

The background emission is taken from "empty" regions in the nearby vicinity. We obtain this by finding the standard deviation of the emission from the closest 50 non-overlapping regions to the object of interest. These empty coordinates are selected from a list of known “empty” regions, where we define these regions as at least 1 arcsecond away from a preexisting source in the segmentation map. This is in order to ensure that these regions do not have the faint outskirts of objects in them. 

We visually inspect the model images of the cutouts for the  quiescent samples produced by \texttt{GALFIT} to classify them. We classify them into 4 categories shown in Table \ref{table:types} - compact or elliptical galaxies (galaxies that are spheroids), disc galaxies, peculiar galaxies that are interpreted as mergers of two preexisting galaxies, and galaxies whose type we cannot determine as they are too faint. We find 77 compact galaxies, 4 disc galaxies, 7 peculiars, and 2 galaxies of unknown types. Results are shown in Table \ref{table:types}.

\subsection{Radius and S\'{e}rsic Index of Quiescent Galaxies}

\begin{table}
    \caption{Number of galaxies in different types after visually examining their morphology. Peculiar galaxies are interpreted as mergers of two preexisting galaxies. Unknowns are galaxies whose type could not be determined as they are too faint.}
    \centering
    \label{table:types}
    \begin{tabular}{p{0.35\linewidth}|p{0.15\linewidth}}
    \hline
    \textbf{Type of galaxy} & \textbf{Number}\\
    \hline
    Compact/Elliptical & $77$ \\
    Disc & $4$ \\
    Peculiars & $7$ \\
    Unknown & $2$ \\
    \hline
    \end{tabular}
\end{table}

Based on the results obtained using \texttt{GALFIT}, we examine the radius and Sérsic index distribution of our sample of galaxies. When looking at these two distributions, we remove all the objects whose radius is smaller than the radius of the PSF itself. The radius of PSF is found to be 2.3 pix in the F444W band \citep{PSF}, so we discard all the galaxies with radii smaller than 2.3 pix. This removes 32 objects from further analysis of radius and Sérsic index. We also remove objects with radii larger than 7 kpc and Sérsic incidices higher than 8 or lower than 0.5 to filter out remove bad fits to the light profile and nonphysical objects. This removes 1 object from our sample. 

Next, we plot the distributions of radius and Sérsic index of the full sample after cutting $10\%$ of the galaxies with the highest RFF values, which indicated that they were the poorest fit. To assess if poorly fit galaxies skew the distributions of morphologies measured from our sample. We conduct a t-test between the Sérsic and radius distributions of the full sample and with the worst $10\%$ of RFF values removed. We find the distributions are consistent with being the same. The average radius and Sérsic index values for our sample are shown in Table \ref{table:hist}. 

\begin{table}
    \renewcommand{\arraystretch}{1.5}
    \caption{Average values, together with their standard deviations, of radius and Sérsic index of quiescent galaxy sample and general galaxy sample (obtained by \citet{Katherine}) at $z>3$.} 
    \centering
    \label{table:hist}
   \begin{tabular}{p{0.3\linewidth}p{0.25\linewidth}p{0.25\linewidth}}
    \hline
    \textbf{} & \textbf{Radius (kpc)} & \textbf{Sérsic index}\\
    \hline
    Quiescent sample  & $1.31 \pm 0.85$ & $1.93 \pm 0.95$\\
    General sample & $1.46 \pm 0.84$ & $1.28 \pm 0.87$\\
        \hline
    \end{tabular}
\end{table}

\section{Results}\label{results}

\subsection{UVJ Colour-Colour Diagram}\label{uvj}

There are several ways in which to identify galaxies which are passive outside of a direct measurement of the star formation rate. One example of this are the so-called UVJ diagrams, which can serve as valuable diagnostic tools in extragalactic astronomy, enabling the relatively accurate determination of whether a galaxy is actively star-forming or quiescent \citep{Patel_2012}. These are constructed by plotting the rest-frame U-V colours of a galaxy against their rest-frame V-J values.  An early version of the UVJ diagram was introduced by \citet{Labbé2005}, in which it was suggested that the diagrams displayed two different kinds of reddened galaxies that could be identified. These were those reddened by age and those by dust.

The ability to discriminate between the effects of age and dust is a very useful tool, as dust frequently causes the misidentification of quenched galaxies. Dust preferentially absorbs blue light, which acts to warm dust grains, causing the reemission of light in the infrared range.  For example, research conducted by \citet{wuyts2007we} demonstrated that galaxies exhibiting red rest-frame U - V colours within the redshift range of $z = 2 - 3.5$ also display redness in the rest-frame V - J, helping the differentiation between young and dusty galaxies versus older quenched ones. 

\begin{figure*}
    \centering
    \includegraphics[width=0.9\textwidth,keepaspectratio]{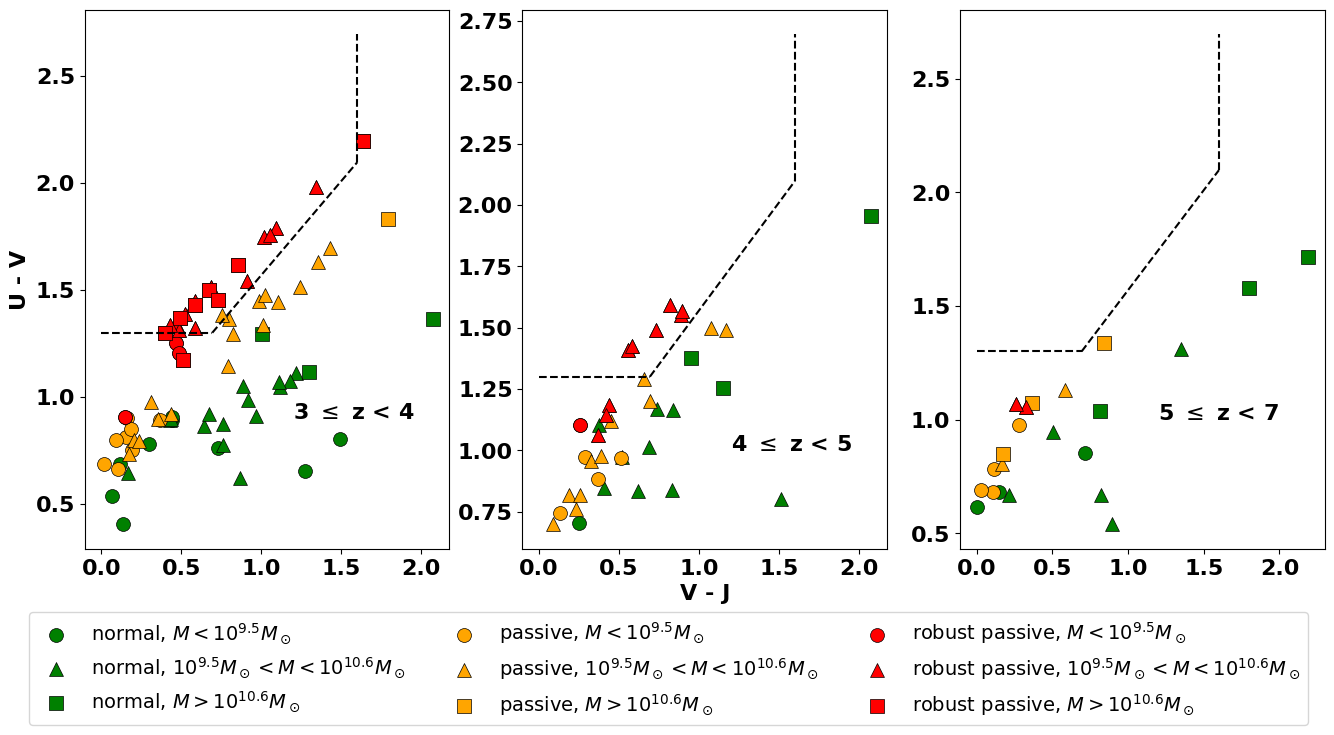}
    \caption{The initial, lenient, \texttt{Le Phare} passive sample which was then run through \texttt{BAGPIPES} to obtain the final passive sample plotted on UVJ diagrams. Galaxies have been split into normal (star-forming), passive and robust passive categories, as well as redshift and mass bins, as indicated in the legend. Where passive and robust passive signify that 50 \% or 97.5\% respectively, of the galaxy's sSFR PDF falls below the threshold set by Equation \ref{eq:sSFR}.
    }
    \label{fig:uvj}
\end{figure*}

In this work, we will utilize the quiescent selection criterion
\begin{equation}
  \label{eq:UVJ1}
  U - V > 0.88 \times (V - J) + 0.69,
\end{equation}
as outlined by \citet{williams2009detection}, where the U - V cutoff is $\sim$ 1.3 and the V - J cutoff is $\sim$ 1.6 for quiescent galaxies with redshifts lower than 0.5.  Generally, the non-dusty and star-forming objects are located to the left of the graph in the low U - V and V - J region, whereas the top right, or high U - V and V - J display the dusty and star-forming objects. Figure \ref{fig:uvj} shows the distribution of our final galaxy sample in colour-colour space, illustrating the principles described above.

Figure \ref{fig:uvj} shows our selection of \texttt{BAGPIPES} fitted galaxies on a UVJ diagram. The colours here highlight the trend of increasingly passive galaxies moving further into the quiescence region described by Equation \ref{eq:UVJ1}. Across mass cuts, we can see that the robustly passive selection tends to be more confidently placed in or near the quiescent region. The “normal”, “passive” and “robust passive” selections appear to be layered on top of one another in a diagonal orientation, which is particularly apparent in the lowest redshift cut. Because of this, we also note that the full passive sample is almost entirely on the outside border of the range of the quiescent section. This trend has also been noticed in other works such as \citet{Valentino_2023} who attempted to combat this by defining a second version of the quiescent UVJ cut which allowed for a more lax definition of quiescence, effectively defining the UVJ colour version of our passive and robust passive cuts. 

We note that as galaxies have more time to quench, they are able to move upward into the quiescence region. This manifests itself in the layering effect mentioned above, where the more recently quenched full passive galaxies have not had the time to move into the selection region. This also causes an effect in higher redshift diagrams, with quenched galaxies increasingly clustered in the bottom region of the graph, which would otherwise be labelled star-forming. Effectively, we note that the higher the redshift the less reliable UVJ methods are at correctly selecting all quenched galaxies. This trend is noted and further elaborated upon in \citet{trussler2024like}, where Figure 3 in that work shows the evolution of a quenching galaxy model on a UVJ diagram over time relative to the classic rest-frame UVJ cut. It is also the case that galaxies at these redshifts obtain their overall morphology as spheroids or ellipticals in appearance before their star formation or colours are completely quenched \citep[e.g.,][]{Conselice2024b}.

Objects frequently become quenched, in terms of star formation, before they have entered the quiescence region, hence we expect an increasingly large number outside of the classical UVJ cut as redshift increases. \citet{trussler2024like} postulates that the SEDs of galaxies at high redshift are likely still being dominated by their recent star formation activity; as such, older methods for selecting passives like the UVJ cut, likely overlook a significant fraction of quiescent galaxies. That is, it takes longer for a galaxy to appear passive in UVJ than it does when examining its star formation \citep[][]{Conselice2024b}. Galaxies which have crossed into the linear extension of this, which usually contains non-dusty star-forming galaxies, are usually overlooked.

To quantify this effect: a galaxy would take $\sim$ 1 Gyr to fully quench and then enter the quiescent region of the UVJ diagram, however at a redshift of z = 5 the universe was approximately $\sim$ 1.2 Gyr old, which would not give objects adequate time to form and then fully move into this cut, as the universe would be too young. Given that passive, or at the least quenching objects, have been discovered at higher redshifts than this, \citep{looser2024recently, weibel2024}, we can confidently conclude that alternative methods, such as those based on the sSFR cut we use, should instead be applied to find these systems. 

Furthermore, at the higher redshift ($z > 3$) the J band component becomes more unreliable as it is extrapolated based on the SED fitting. This is due to the rest-frame J band shifting beyond the coverage of NIRCam and hence it cannot be directly observed. This again is makes the UVJ selection more unreliable at the higher redshifts. 

\subsection{Co-moving Number Density Evolution and Evolution of Quiescent Galaxy Fraction}

\begin{figure}
    \centering
    \resizebox{0.45\textwidth}{!}{\includegraphics{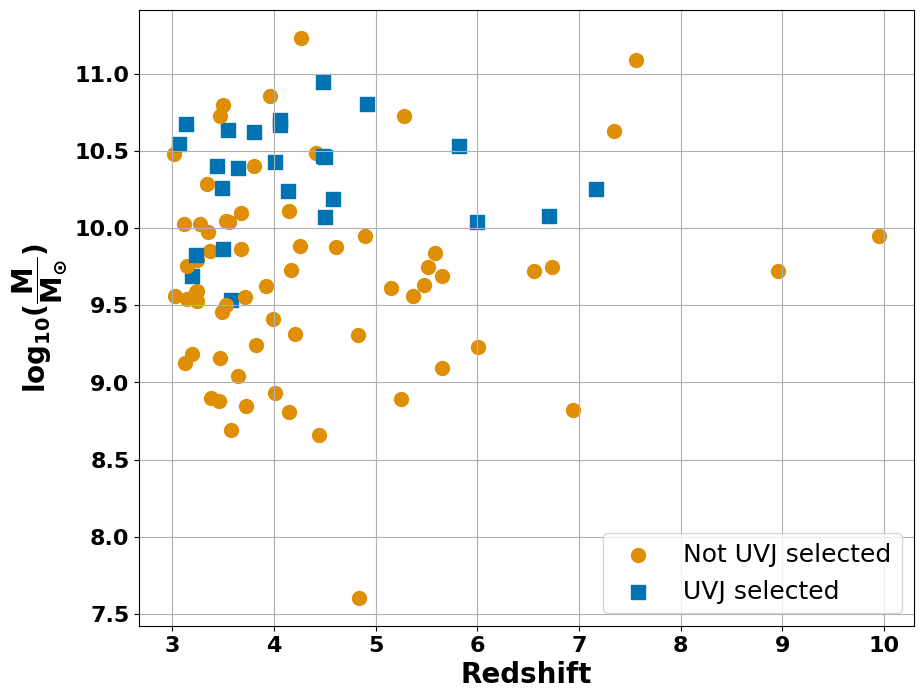}}
    \caption{Redshift versus mass diagram for the passive sample defined by our sSFR cut in Equation \ref{eq:sSFR}. The orange dots indicate the quiescent galaxies that have not been selected via UVJ selection. The blue squared indicate the quiescent galaxies that have been selected via UVJ selection.
    }
    \label{fig:mass_z}
\end{figure}

Before we start with the analysis of co-moving number density evolution and evolution of quiescent galaxy fraction, we analyse the dependence of mass of the observed candidates with redshift. In Figure \ref{fig:mass_z} we plot the redshift versus stellar mass diagram to establish some trends. On the plot, we mark quiescent galaxies that are also selected with the UVJ selection. From the plot, we can see that the low mass galaxies ($\log_{10}{(M_{\star}/M_{\odot})} \leq$ 9) are not common. This is due to completeness, which we estimate to rapidly fall off below 10$^9$ in our simulation work (see Section \ref{compl}).

We also note that the stellar mass of the galaxies increases with increasing redshift, which is due to the detection limits at higher redshifts preventing the detection of lower mass and therefore fainter galaxies.

\begin{figure*}
    \centering
    \includegraphics[width=0.8\textwidth,keepaspectratio]{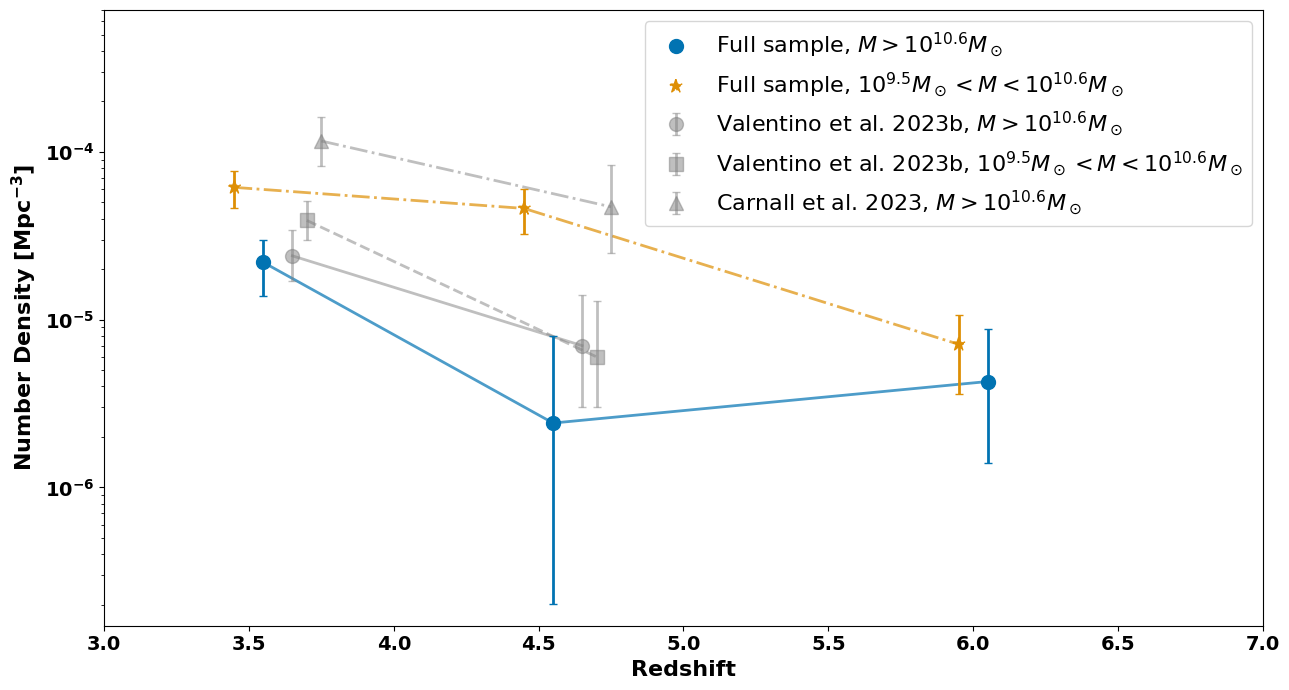}
    \caption{The evolution of co-moving number density of massive quenched galaxies across redshift and mass cuts for our combined fields. The x-axis values are averages of the redshift bin intervals they span. The uncertainties in the figure
    incorporate cosmic variance as well as the Poisson noise. We also include values from previous work, \citet{Valentino_2023, carnall2023}, for comparison.
    }
    \label{fig:evo1}
\end{figure*}

\begin{figure*}
    \centering
    \includegraphics[width=0.8\textwidth,keepaspectratio]{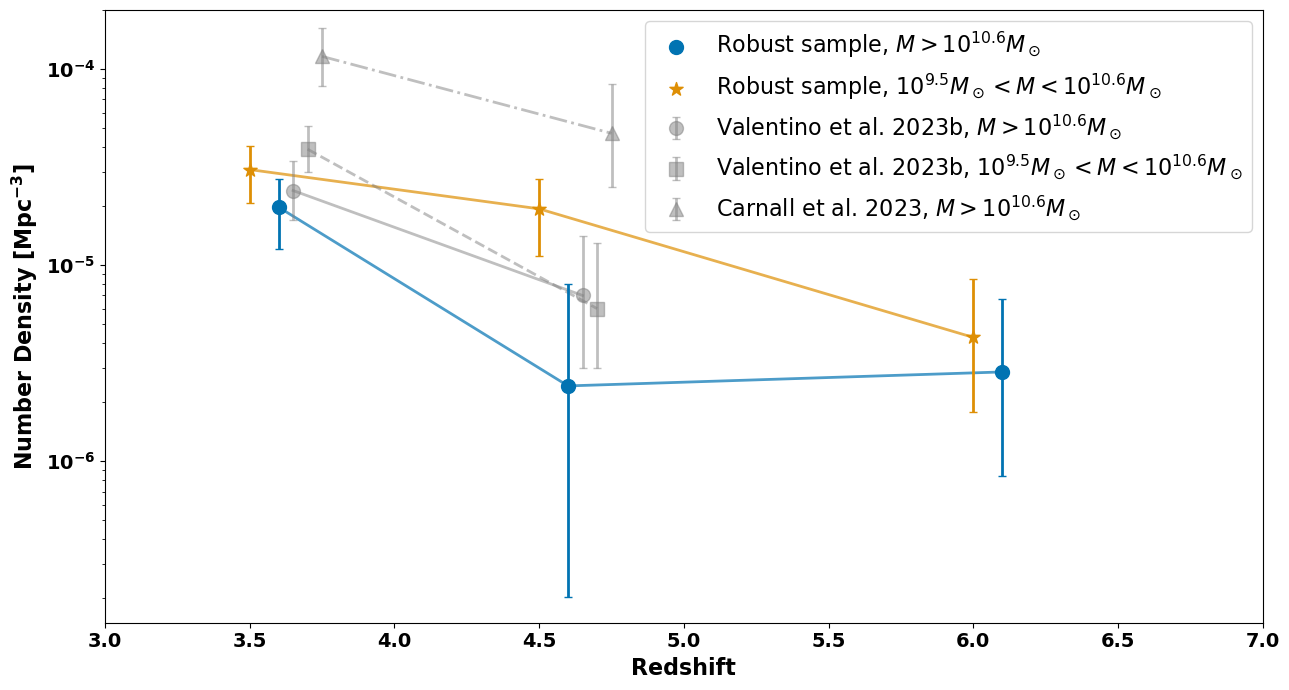}
    \caption{The evolution of comoving number density of massive robustly quenched galaxies across redshift and stellar mass cuts for our combined fields. The x-axis values are averages of the redshift bin intervals they span. The uncertainties in the figure
    incorporate cosmic variance as well as the Poisson noise. We also include values from previous work, \citet{Valentino_2023, carnall2023}, for comparison.}
    \label{fig:evo2}
\end{figure*}

We perform a detailed analysis of the number densities for different stellar mass samples across redshift in this section. Figures \ref{fig:evo1} and \ref{fig:evo2} show how number density evolves with redshift for the full and robust passive samples respectively, with additional data from \citet{Valentino_2023} and \citet{carnall2023} added for comparison.

In this analysis, we include only the high and medium-mass samples. While we detect galaxies in our low-mass sample, we do not show their number densities as the sample is likely incomplete, and there is no obvious robust way to correct for this incompleteness.

The process of calculating field volumes for the number densities took into account the geometry of the individual fields. We converted the total unmasked area of each field into a volume in $Mpc^{3}$ using the co-moving distances for the limiting values of the redshift bins. We also combine all three fields, giving a total area of 121.47 $arcmin^{2}$, and find the overall co-moving number density of these fields using the sSFR-motivated method of identifying massive quiescent galaxies.

Our observations find an overall declining number density of passive galaxies as a function of increasing redshift, in broad agreement with the \citet{Valentino_2023} and \citet{carnall2023} data. We find the robust passive number density is consistently lower than that of the full passive sample. However, the relative rarity of passive systems, small volumes probed by JWST, and decreasing completeness at higher redshifts currently limits any conclusions above redshift z=5. 

All samples show consistent results at the highest redshift cut, however the highest mass curves deal with the issue of inherent rarity. High-mass galaxies are inherently rare, meaning that current field depths are not a limiting factor, and so this sets a finite limit on the number detectable in a given area of observation.

We note that our results show the most overall consistency with the two samples from \citet{Valentino_2023}, with a greater agreement between these and the full passive sample than with the robust passive. We do not observe the same steep decline as \citet{Valentino_2023} in our middle mass cut, and we also report a much higher density overall across redshift within this mass range. As \citet{Valentino_2023} uses the UVJ method to identify passive candidates, we expect disagreement at the higher redshifts, as explained in Section \ref{uvj}. Although the sample from \citet{carnall2023} covers our high mass range, we find the best agreement between it and our middle mass selection. It is worth noting that \citet{carnall2023} only uses 4 NIRCam paintings from CEERS which appear to be overdense. For further comparison to this work, see Section \ref{comparingcarnall}
As there is a limit to the redshift range covered by these works, further comparison is limited. 

\begin{figure*}
    \centering
    \includegraphics[width=0.7\textwidth,keepaspectratio]{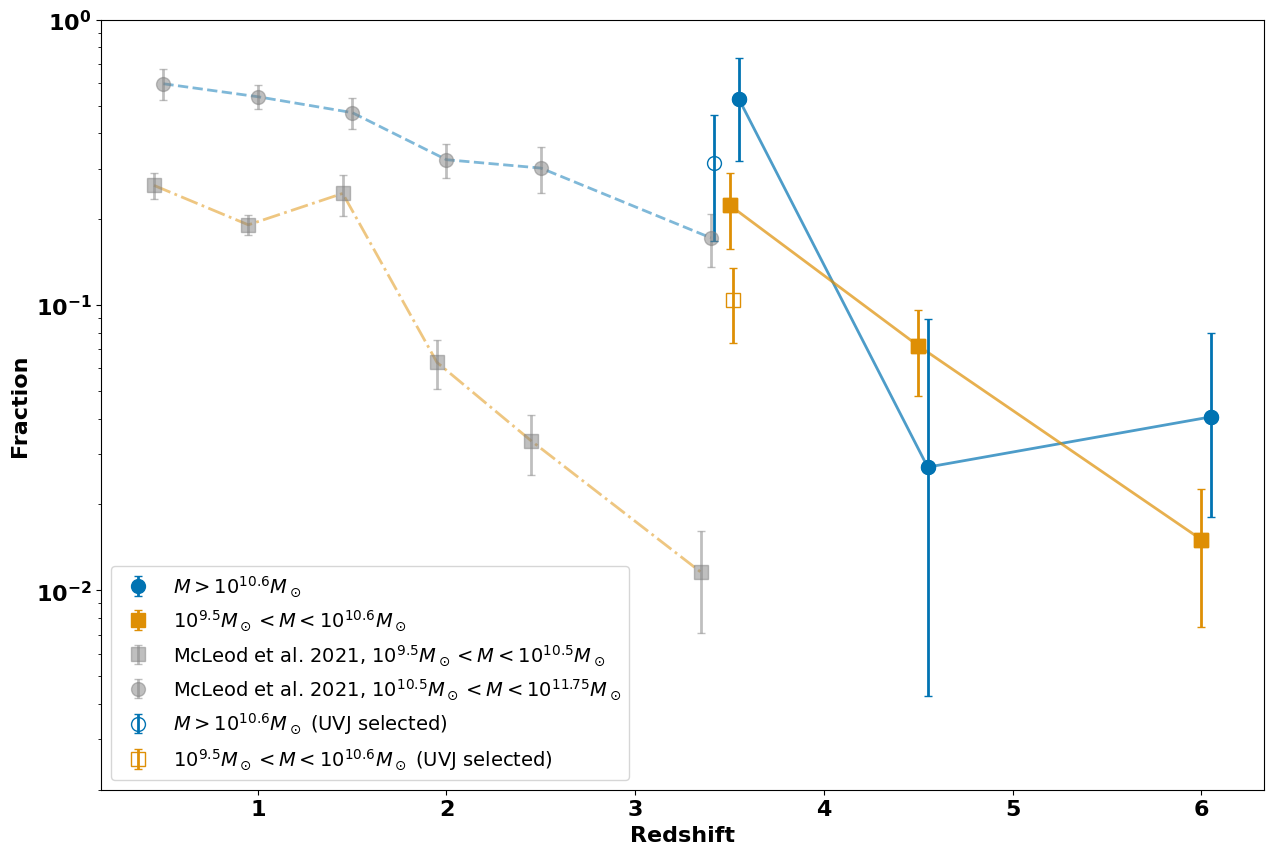}
    \caption{The evolution of the fraction of massive quenched galaxies to total galaxies found across redshift and stellar mass cuts for our combined fields using our full passive sample. The x-axis values are averaged of the redshift bin intervals they span. Data from \citet{McLeod2021} at lower redshifts is included for comparison. The hollow data points indicate the fractional values for our UVJ selected quiescent sample, and are also included for comparison. Note that the values from this work span the redshift bins $3 \leq z < 4$, $4 \leq z < 5$, and $5 \leq z < 7$.
    }
    \label{fig:fract}
\end{figure*}

Number densities inform us about the absolute numbers of passive galaxies. To contextualise them with the wider galaxy population, we estimate the fraction of quenched galaxies in each mass bin (Figure \ref{fig:fract}). We again include only high and medium-mass bins in this analysis but leave out the low-mass bin due to a completeness issue. The uncertainties in the figure incorporate the Poisson noise from the number counts of galaxies, as the cosmic variance over the total area is sub-dominant to the poison error. We find the trend between data points for the medium-mass bin can be described by a power law slope of the form $f(z) = A \times (1 + z)^{n}$, where $A$ is a normalising constant and $n$ is an indicator of how rapid the evolution is. We find this power-law slope is best described by the values A = 2.03 $\pm$ 0.86 $\times$ 10$^{3}$ and n = -6.04 $\pm$ 0.25.

We also include data from \citet{McLeod2021} in this figure for comparison and to build a picture of passive fraction behaviour across a wider redshift range. \citet{McLeod2021} observe a fairly consistent fraction at lower redshifts for the highest mass cut, with a slow decline at higher redshifts. We observe a higher quenched fraction between $z = 3 - 4$, followed by a significant drop between $z = 4 - 5$, and then a slight increase in the final redshift bin. This drop causes the power law fit to continue to cut off steeply, being unable to capture the behaviour in the final redshift bin. The overall behaviour of the highest mass cut shows some level of consistency with the \citet{McLeod2021} data, however, our errors are too large to be sure. 

While we observe a fairly smooth decline for the medium-mass sample, which is reflected in the accuracy of our power law fit for this mass cut, there is a significant gap in our sample and that of \citet{McLeod2021}. This is in part due to the different methods for isolating passives used. \citet{McLeod2021} uses the same UVJ identification criterion as \citet{Carnall2017} and  \citet{Carnall2020}. Using a different criterion will create inconsistency, which as mentioned also has issues at higher redshifts, which would present more of an issue if the results in \citet{McLeod2021} extended further. To provide a more direct comparison with our data, we also include data points from our UVJ selected sample in the redshift 3 - 4 range. The two UVJ selected samples show less of a discrepancy than the \citet{McLeod2021} sample and the sSFR selected one. We further compare our results when using the UVJ selection to find quiescent galaxies in Figure \ref{fig:uvj}.  As can be seen, we find using the UVJ method of identification reduces our sample by a large factor, showing that UVJ finds not only quenched galaxies, but those that have time to evolve into systems dominated by older stellar populations. 

Some of our results are limited by the low number of galaxies in our sample, this is again, due to completeness issues and the inherent rarity of higher-mass objects. Overall, we observe high fractions of passive galaxies, indicating that both the total galaxy number and passive number must increase at similar rates across redshift. The high presence of quenched galaxies relative to overall galaxy numbers in the early universe could have important implications for our understanding of galaxy evolution.

\subsection{Comparisons to Previous Work}

\begin{figure*}
    \centering
    \includegraphics[width=0.7\textwidth,keepaspectratio]{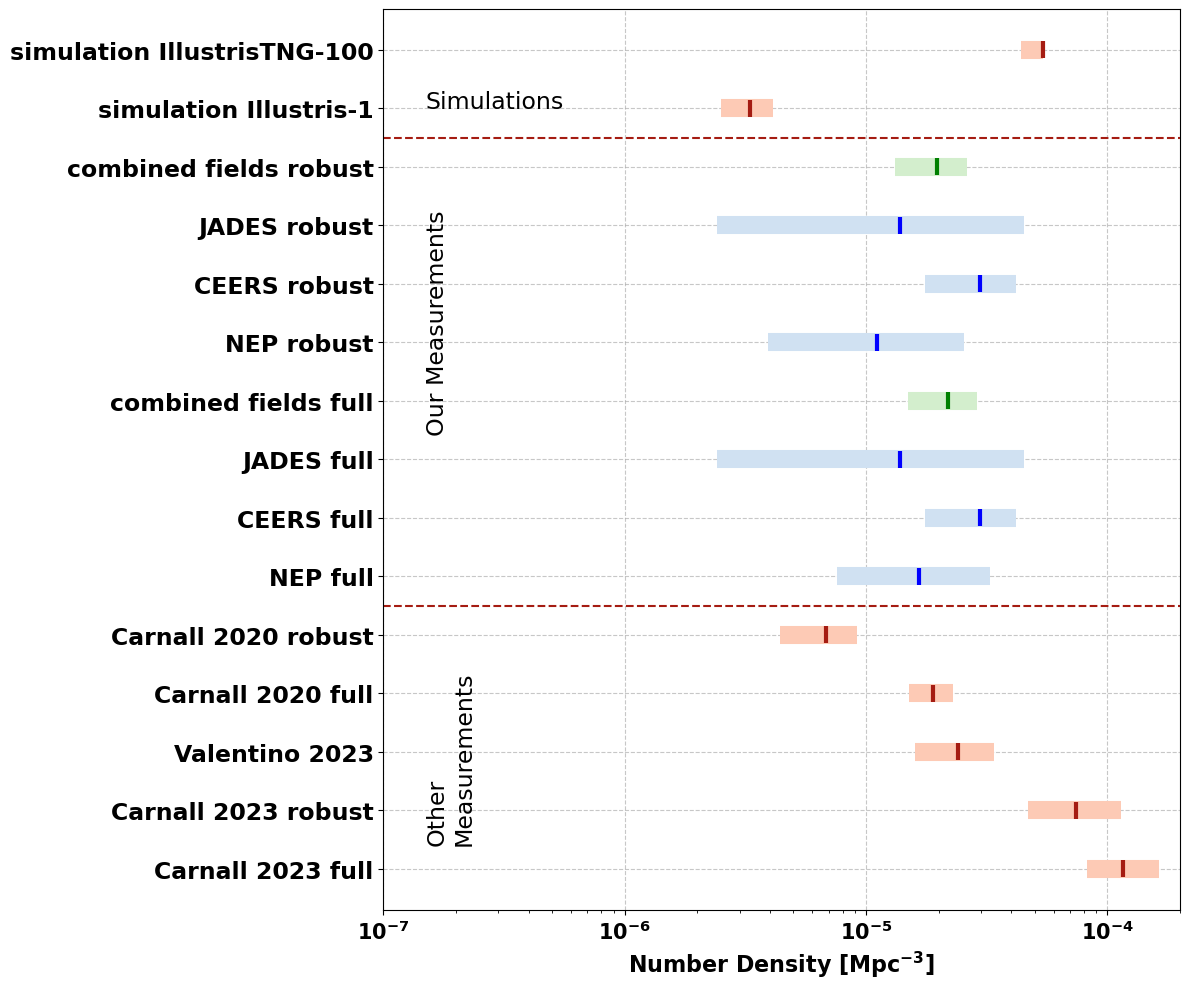}
    \caption{The comoving number densities of massive quiescent galaxies reported in this work and from literature. The values cover the redshift interval, 3 $\leq$ z $<$ 4, and the higher mass cut, $M > 10^{10.6} M_{\odot}$. The uncertainties here do not include the contributions of cosmic variance, and instead only display the Poisson noise. This was done for consistency with comparisons from the literature, such as in \citet{Valentino_2023}, who do not include cosmic variance. The results from $Illustris$-1 and $Illustris$-TNG-100 have number densities only at z = 3, instead of the redshift range 3 $\leq$ z $<$ 4.}
    \label{fig:numdensmainplot}
\end{figure*}

Following the argument outlined in \citet{Valentino_2023}, we attempt to provide a meaningful comparison between the results of different works by contrasting comoving number densities over the same mass and redshift interval. As is reported in Figure \ref{fig:numdensmainplot}, we compare against results using the most commonly studied mass and redshift bin (3 $\leq$ z $<$ 4, $\log_{10}{(M_{\star}/M_{\odot})} \geq$ 10.6).

We make use of the simulational results from $Illustris$-1 \citep{Illustris} and $Illustris$-TNG-100 \citep{nelson2021illustristng}, as well as survey results from \citet{Carnall2020}, \citet{carnall2023} and \citet{Valentino_2023}.
The simulation results from $Illustris$-1 and $Illustris$-TNG-100 provide number densities for redshift $z = 3$, instead of for the redshift range $3 < z < 4$. This constraint needs to be taken into account when directly comparing the densities. We use the number density calculation similar to that made in \citet{Valentino_2023}, wherein they take data from both simulations and apply a sSFR cut by filtering only for galaxies where the sSFR $\times$ 10$^{−10}$ yr$^{−1}$ within 2 $\times$ the half-mass radius.  This best matches how we select our galaxies within the data itself and thus allows for a direct comparison.

Our findings show the strongest agreement with the results from \citet{Carnall2020} and \citet{Valentino_2023}. The latter work uses the UVJ criteria described by Equation \ref{eq:UVJ1}, whereas the former uses the same sSFR-based cut as this work, described by Equation \ref{eq:sSFR}. This consistency between results shows promise for either method to be used interchangeably to identify quiescent samples at redshifts z $\sim$ 3. The primary result of this analysis, the full sample combined field number density, is of the order of $\sim 2.1 \times 10^{-5} Mpc^{-3}$ and is best concretely compared with these. We find that this value shows the best agreement with the full sample from \citet{Carnall2020}, being $\sim 1.1 \times$ larger, as well as the strict UVJ sample from \citet{Valentino_2023}, being $\sim 1.1 \times$ smaller than this. Our estimate is $\sim 3 \times$ larger than the robust selection reported by \citet{Carnall2020}. The other estimates we report span the range of all results barring the number densities from $Illustis$-TNG-100 and \citet{carnall2023}. We expect their results to indicate a slightly higher number density than would otherwise be produced. We find that $Illustris$-1 produces a notably lower comoving number density than that of $Illustris$-TNG-100. This is to be expected as the latter is an updated version of the same simulation, which incorporates updated treatment of AGN and stellar-wind feedback. The combined result is much smaller than the results from \citet{carnall2023}; this is examined further in Section \ref{comparingcarnall}. 

\subsection{Multiwavelength \& Ancillary data of Quiescent galaxies}\label{multiwave}

\begin{table*}
\caption{Quiescent galaxies found in the other regions of the spectrum. The Photometric $z$ column indicates the redshift of the object found by \texttt{BAGPIPES}. The MIR, X-ray and Radio columns indicate whether the object was found in the mid-infrared catalogue, X-ray or radio catalogue, respectively. The Radio Luminosity column indicates the radio luminosity of the object found in the radio catalogue, where the frequencies are 1.4 Ghz, 3 Ghz and 1.4 Ghz for the CEERS, NEP and JADES fields respectively.}

\centering
\label{table:multiwave}
\begin{tabular}{cccccccccc}

\hline
FIELD & ID    & Photometric $z$ & MIR & X-ray & Radio & Radio Luminosity & X-ray Luminosity \\
 & & & & & & $W \cdot 10^{24}$ & erg s$^{-1}$ cm$^{-2} \cdot 10^{-16}$\\
\hline
CEERS & 12017 & $3.15_{-0.12}^{+0.13}$ & no                & yes   & no    & /  & $5.98_{-2.10}^{+2.30}$            \\
CEERS & 18901 & $3.13_{-0.10}^{+0.10}$ & yes             & no    & no    & /  & /              \\
CEERS & 25071 & $3.71_{-0.32}^{+0.20}$ & yes               & no    & no    & /  & /              \\
CEERS & 26564 & $3.48_{-0.25}^{+0.27}$ & no              & yes   & no    & /  & $7.47_{-1.47}^{+1.74}$              \\
CEERS & 27157 & $3.72_{-2.58}^{+0.40}$ & no               & yes   & no    & /  & $6.43_{-1.46}^{+1.75}$              \\
CEERS & 43098 & $8.99_{-0.23}^{+0.24}$ & yes              & no    & no    & /  & /              \\
CEERS & 46917 & $6.01_{-0.20}^{+0.21}$ & yes         & no    & no    & /  & /              \\
CEERS & 74351 & $4.96_{-0.15}^{+0.14}$ & yes            & no    & no    & /  & /              \\
NEP   & 46844 & $3.96_{-0.34}^{+0.18}$ & no        & no    & yes   & $11.1_{-2.3}^{+1.3}$ & / \\
NEP   & 54448 & $3.25_{-0.10}^{+0.11}$ & no       & no    & yes   & $1.13_{-0.21}^{+0.21}$ & / \\
NEP   & 72863 & $4.37_{-0.13}^{+0.13}$ & no     & no    & yes   & $1.95_{-0.35}^{+0.35}$ & / \\
JADES & 30450 & $3.61_{-0.18}^{+0.17}$ & no               & yes   & no    & /  & $3.53_{-0.34}^{+0.38}$              \\
\hline
\end{tabular}
\end{table*}

We examine the data from other ranges of the electromagnetic spectrum to see if our quiescent sample has been detected. Detection at other wavelengths is of interest, as it may help reveal the presence of AGN activity. We first examine publicly reduced MIR data from MIRI taken by the CEERS team \citep{yang2023ceers}. We find a total of 8 cross-matches and for those remaining in our sample, their specifications are given in Table \ref{table:multiwave}. To assess what impact the inclusion of MIRI data has on the selection of our passive galaxies, we repeat our \texttt{Le PHARE} SED fitting with available MIRI photometry from \citep{yang2023ceers}. We find that 4/8 sources have negligibly small changes to their physical properties ($<0.1$ dex). Two sources have  $\sim0.5$dex more star formation and two sources have more than 1 dex increase in SFR. These two sources with a large increase in SFR are better fit with significantly more dust (0.25-0.6 mags extra extinction). For all MIRI-detected sources, stellar masses are stable within 0.15 dex. A total of 3 sources (only 1 robust) would no longer pass our sSFR cuts with MIRI data included, indicating there is some contamination from dusty galaxies entering our sample. 

Secondly, we checked our data against the 0.5 – 7 keV X-ray Chandra observations \citep{nandra2015aegis}, the 0.5 - 10 keV Chandra data \citep{luo2016chandra}, and the 0.5 – 10 keV XMM-Newton data \citep{zhao2024pearls}, covering the CEERS, JADES, and NEP fields respectively. In the CEERS, field we find 3 matches with our quiescent sample. In the JADES field, we find only 1 match, while in the NEP field, we find no matches. Matches found are given in Table \ref{table:multiwave}. This brings the total number of quiescent galaxies identified in the X-ray catalogue across all three fields to 4, meaning only $4.4\%$ of the objects we identified as quiescent in all the fields were found in X-ray catalogues as well.  

As AGNs all emit X-rays \citep{Haardt1991, Elvis1978}, the detection of X-ray emission in a galaxy is often a sign of AGN activity. The presence of an AGN in a passive galaxy inferred through its X-ray emission could be an indication of the AGN acting as the main driver of quenching for these galaxies. 

We checked our data against radio catalogues using SCUBA-2 data from the JCMT, as well as 3 GHz VLA data. Both the JWST and VLA have much smaller PSFs than SCUBA-2, hence we focus on using VLA catalogues to limit confusion between different sources. In both the CEERS and JADES fields, we find no matches with our quiescent sample. In the SCUBA Survey of the NEP field, we find 1 match. In the VLA survey of the NEP field, we find 3 matches, with 1 match being found both in the VLA and SCUBA survey. Matches found are given in Table \ref{table:multiwave}. This brings the total number of quiescent galaxies identified in the radio catalogue across all three fields to 3, meaning only $3.3\%$ of the objects we identified to be quiescent in all the fields were found in radio catalogues. 

Since the identification of galaxies in radio data can be either a sign of AGN or active star formation, we calculate their radio luminosity. Their calculated luminosities are found in Table \ref{table:multiwave}. High radio luminosities usually indicate AGNs, while lower luminosities indicate active star formation. Based on the work done by \citet{Smol_i__2017} and \citet{Novak_2018}, luminosities of the order $\approx 10^{24}$W, can be indicative of either AGN or star formation activity. However, when combined with our rest-frame UV and optical analysis with BAGPIPES, the radio emission is likely the result of an AGN as opposed to star formation. If we look at the low dust attenuation estimated from \texttt{BAGPIPES} for these galaxies, we see that it unlikely for these galaxies to be actively forming stars as they would have to be extremely dust-reddened. We conclude that these galaxies contain AGNs, which are possibly the main driver of the quenching in these galaxies.

\begin{figure}
    \centering
    \resizebox{0.5\textwidth}{!}{\includegraphics{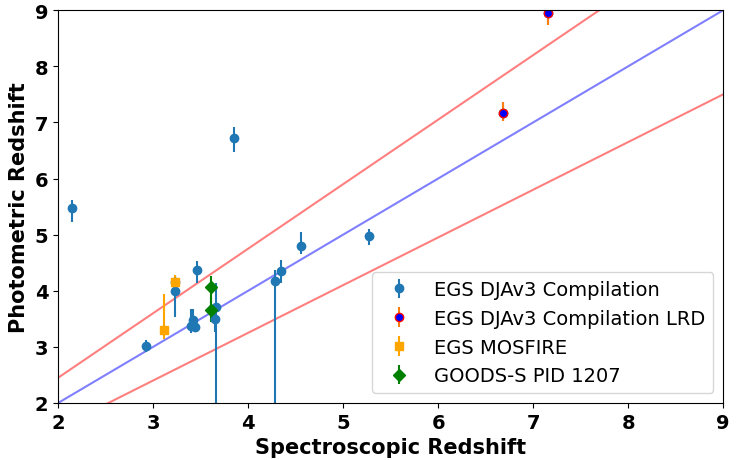}}
    \caption{Photometric redshifts fit from our sample vs 
spectroscopic redshifts fit using data from several catalogues. We indicate where values match and where the 15 percent error boundary in $1+z$ is with blue and red lines, respectively. We also specifically indicate two objects found to match known LRDs.
We use data from the the 3D-HST Survey \citep{CEERS_spec1}, the MOSFIRE Deep Evolution Field (MOSDEF) survey \citep{CEERS_spec_2} and the v3 release from the Dawn JWST Archive (DJA) \citep{heintz_2024}}
    \label{fig:spec_z}
\end{figure}

Thirdly, we also look at the spectroscopic catalogues. Figure \ref{fig:spec_z} shows a comparison between the photometric redshifts fit from our sample and 
spectroscopic redshifts fit using data from several catalogues. We indicate where values match and where the 20 percent error boundary is with blue and red lines, respectively. We begin by examining some non-JWST sources of spectroscopic redshifts. In the 3D-HST Survey \citep{CEERS_spec1} of the CEERS field, we find 3 matches. In the MOSFIRE Deep Evolution Field (MOSDEF) survey \citep{CEERS_spec_2} we find 4 matches, with 1 match being found both in the MOSFIRE and 3D-HST Survey. Of these 6 unique sources, only two sources have secure redshifts, with one in close agreement with our photo-z and one approximately 20\% away. The CEERS field has had coverage from JWST NIRSpec from multiple survey programmes. We compare our photometric redshifts to the v3 release from the Dawn JWST Archive (DJA) \citep{heintz_2024}, which contains the original CEERS JWST spectroscopy as well as recent programmes such as RUBIES \citep{DeGraaf2024}. We find a total of 17 cross-matches, with 11 targets having consistent spectroscopic redshift measurements (within 15\%) and a further four lying just on the border of this classification ($\sim20$\% deviation) and 2 extreme outliers with $\Delta z \sim 3$. These spectra originate from the following observing programmes: CEERS \citep[PID 1345][]{Bagley_2023}, CEERS DDT \citep[PID 2750][]{ArrabalHaro2023}, NIRSpec WIDE MOS Survey \citep[PID 1213][]{Maseda2024}, a passive galaxy focused study \citep[PID 2565][]{Nanayakkara2024} and RUBIES \citep[PID 4233][]{DeGraaf2024}. Within the JADES field, there are an additional 2 spectroscopic cross-matches which are both from PID 1207 \citep{Rieke2024}. Both of these sources have closely matching photo-z's. In all cases where photometric redshifts disagreed by more than 15\%, the photo-z was higher than the spectroscopic redshift. The two highest redshift sources with a spectroscopic redshift above 6 are classed as candidate `Little Red Dots' (LRD). These are CEERS-45647/RUBIES-EGS-49140 and CEERS-43098/RUBIES-EGS-42803 \citep[See also][]{Wang2024}. We discuss such sources at the highest redshifts in Section \ref{High z}.

Previously, limitations in the mass confirmation of passive galaxy redshifts at $z > 3$ from the ground have hindered spectroscopic studies. With the advent of JWST, there is a strong drive to study passive objects spectroscopically. There have been recent proposals accepted which seek to look at this, such as the Early eXtragalactic Continuum and Emission Line Science (EXCELS) \citep{Carnall_2023exc}. This survey has already produced significant results, such as evidence towards settling a debate on whether certain ultra-massive quiescent galaxies violate $\Lambda$-CDM cosmology, concluding that they do not, but they do suggest extreme galaxy formation physics during the 1st billion years of the universe \citep{carnall2024jwst}. The low number of spectroscopically matched objects was anticipated, as passive galaxies often do not have strong emission line signals. As spectroscopy relies on these, being unable to detect them serves to validate our expectations for passive galaxy data in this way. Passive galaxy detection normally requires deeper spectra in order to detect spectral lines in absorption instead of emission.

\subsection{Radius and Sérsic index distribution of quiescent galaxies}

\begin{figure}
    \centering
    \resizebox{0.5\textwidth}{!}{\includegraphics{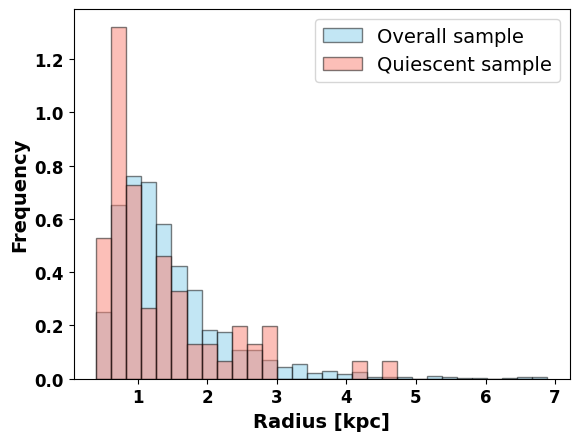}}
    \caption{Normalised radius distribution. Red indicates our quiescent sample. Blue indicates general sample at $z > 3$ obtained by \citet{Katherine}.}
    \label{fig:radius_hist}
\end{figure}

\begin{figure}
    \centering
    \resizebox{0.5\textwidth}{!}{\includegraphics{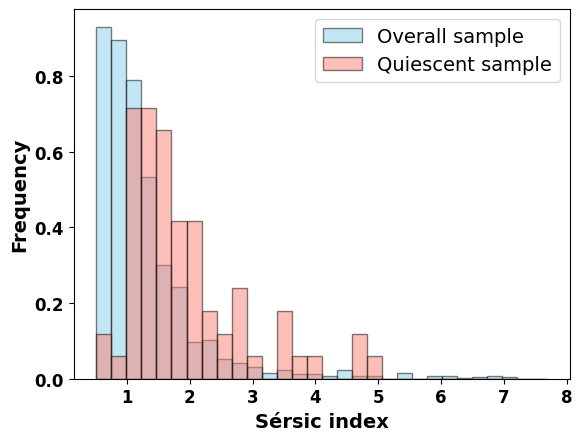}}
    \caption{Normalised Sérsic index distribution. Red indicates our quiescent sample. Blue indicates general sample at $z > 3$ obtained by \citet{Katherine}.}
    \label{fig:sersic_histogram}
\end{figure}

We compare the size and the Sérsic index of our sample to the general sample obtained by \citet{Katherine}. \citet{Katherine} includes a size and structural analysis of 1395 galaxies at $0.5 < z < 8$ within the JWST Public CEERS field that overlaps with the HST CANDELS EGS observations. \texttt{GALFIT} was used to obtain Sérsic models of the rest-frame optical profile of galaxies. We cut the data obtained by \citet{Katherine} so we include only galaxies at redshifts $z > 3$.

Figure \ref{fig:radius_hist} shows the normalised histogram of the radius distribution of our sample together with the histogram of the general sample obtained by \citet{Katherine}, which has been matched to our sample in stellar mass. We perform the t-test and find a P-value of 0.136, indicating that there is no statistically significant difference between our quiescent sample and the general sample obtained by \citet{Katherine}. The average radii values for both samples are shown in Table \ref{table:hist}. 

Next, we divide both our sample and the sample from \citet{Katherine} into mass bins (the same high mass, medium mass, and low mass bins defined previously). We find that there is a significant difference when comparing distributions in the high mass bin, but no significant difference when comparing distributions in the medium mass and low mass bin. Our data indicates that the quiescent galaxies in the high-mass bins are significantly smaller than the galaxies of the general sample in the high-mass bin. 

The average Sérsic index values for both samples are shown in Table \ref{table:hist}. Figure \ref{fig:sersic_histogram} shows the normalised histogram of the Sérsic index distribution of our sample together with the histogram of the general sample obtained by \citet{Katherine}. We again perform the t-test and find a P-value of
1.49 $\times$ $10^{-9}$, indicating that there is a statistically significant difference between our quiescent sample and the general sample obtained by \citet{Katherine}. We find a significantly higher Sérsic index in the quiescent sample than in the general sample, which supports the hypothesis that quiescent galaxies are much more compact.  

We compare the Sérsic index values of our sample and those from the sample from \citet{Katherine}, and find that there is no significant difference when comparing distributions in the high mass bin, but there is a significant difference when comparing distributions in the medium and low mass bins. The low number count of high-mass galaxies likely influenced why we observed no significant difference in this mass bin.

\subsection{Morphology of early quiescent galaxies}

Based on the results reported in Table \ref{table:types}, we conclude that the quiescent objects we find are predominantly elliptical or compact objects, as expected for this galaxy population. We observe very few spiral or disk-like galaxies, which is mostly consistent with the belief that disc-like galaxies evolve into elliptical objects as they merge with other galaxies and quench and use up gas reserves. We find more peculiar than spirals; however, these are still incredibly rare. We expect this as peculiars inherently require a specific time-frame in a rare event, such as galaxy merges, in order to form and be visible. 

\begin{figure*}
    \centering
    \includegraphics[width=1\textwidth,keepaspectratio]{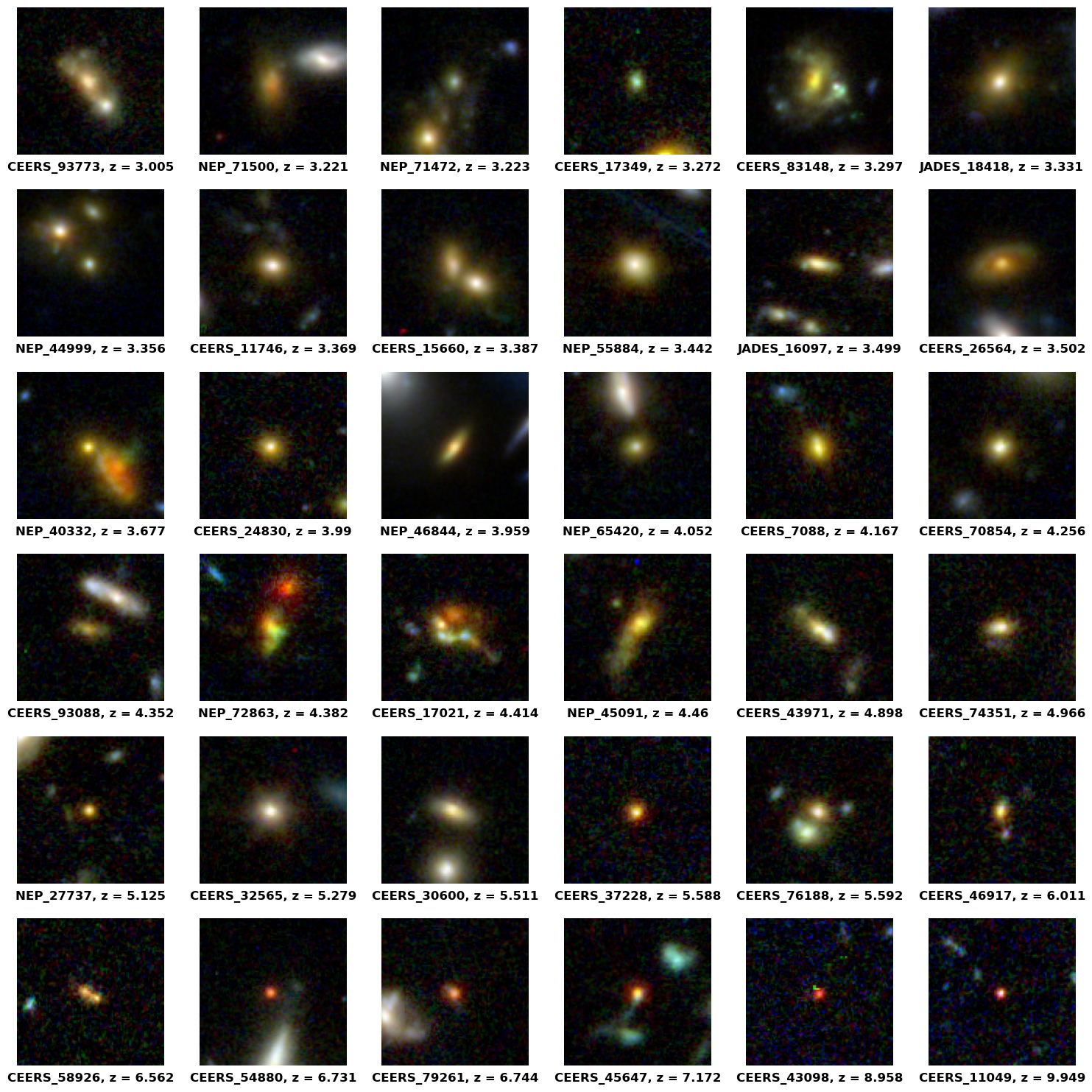}
    \caption{A mosaic of 6 by 6 images ranging from z $\sim$ 3 to z $\sim$ 10. Each image is made into a 101-by-101 pixel size, which corresponds to a $\sim$ 6 arcsecond diameter. The mosaic highlights more interesting and abnormal passives than are proportionally found in our sample, with the majority of objects found to be compact ellipsoids, such as object 24830 in the CEERS field.}
    \label{fig:mosaic}
\end{figure*}

Figure \ref{fig:mosaic} shows a mosaic of galaxies across the redshift range of z $\sim$ 3 - 10. Here we present a selection of quiescent objects passing our selection cuts. It is important to note that the vast majority of objects found are small compact ellipsoids. We display more "abnormal" objects than are proportionally found in our sample, to highlight particularly interesting objects. This is done in the interest of better understanding these objects, their characteristics, and their development. These include diffuse objects, disk-like objects, mergers and potential mergers, "red dots", galaxy groups and pairs, and otherwise strange features. An example of such an object is CEERS-43971, which appears to be a diffuse ellipsoid.  JADES-16097 is an example of a disk-like structure, whereas NEP-46844 appear to be an extended elliptical or a disk-like object.   These objects, despite their structure, are all identified as low star-forming systems. 

During our inspection, we also find several galaxies that are in pairs. Some examples include quiescent galaxies 15660 and 15790 in the CEERS field (which are paried together), 30600, 76188 in the CEERS field, as well as objects 40332, 71500, and 72863 in the NEP field, whose companion objects are star-forming galaxies. In these examples, there appears to be an exchange of some material between the galaxies. Moreover, both of the merging galaxies are observed in the same cutout picture, of size 3 \arcsec $\times$ 3 \arcsec. Their photometric redshifts are also found to be very similar, and hence we can conclude that these are merging galaxies and not galaxies that happen to be chance superpositions.  

We locate some late-phase galaxy mergers as well. Notable examples include galaxies 83148 in the CEERS field and 34464 in the JADES field.  There are also some galaxies that appear to be in the process of forming elliptical galaxies from peculiars, such as 45091 and 58270 objects in the NEP field.   One of the galaxies in the CEERS field (45647),  at redshift $z = 7.17$, is a "little red dot". These objects are red point-like sources whose origin is unknown but are theorised to be compact galaxies with active galactic nuclei, which have been obscured and thus reddened by dust \citep{kokorev2024census}. Other theories also point to these as a link between postulated supermassive black hole seeds in the early universe and observed "problematic" blue quasars. \citep{Kokorev2024, matthee2024little}. 

Evidence, such as \citet{Buitrago2013} and \citet{bell2012turns}, indicates a strong link between galaxies displaying increased S\'{e}rsic index values, reflecting steeper surface brightness profiles, and the quenching of star formation. \citet{bell2012turns} finds that the vast majority of quiescent galaxies have high S\'{e}rsic indices. They examine the correlation between quiescence and several variables, and find S\'{e}rsic index to correlate the best over all redshifts observed. \citet{bell2012turns} and \citet{Fan_2013} also observe this trend to correlate across all redshifts observed.

This has also been seen within simulations by \citet{Tacchella2016}, wherein they confirm that their simulated galaxies gain mass as they quench, along with higher S\'{e}rsic index values. A study by \citet{Dimauro2019} also demonstrates that there exists a direct connection between the rise in central stellar density and the expansion of the bulge component within galaxies. Consequently, the existence of a noticeable bulge in galaxies coincides with their rate of star formation. 

In the nearby universe we typically find that spiral galaxies are younger and larger in radius and have smaller S\'{e}rsic indices of $n \sim 1$, showing their low degree of central brightness concentration. Elliptical galaxies are generally older, with smaller radii and higher S\'{e}rsic indices of $n \sim 4$, the de Vaucouleurs profile, or higher. However, this universality does not hold true across cosmic history, and breaks down at higher redshifts \citep[e.g.,][]{Conselice2011, Conselice2024b}. Observations at high redshifts demonstrate that distant galaxies manifest distinct properties and features compared to their counterparts with the same morphologies in the nearby universe. Notably, studies have revealed variations in the S\'{e}rsic index, colour, levels of star formation, and size among galaxies of the same type across different epochs \citep{conselice2011tumultuous, mortlock2013redshift}. 


There are studies showing that the galaxy morphology depends not only on the star formation rate of a galaxy but on its environment as well. Studies by \citet{Goto_2002} and \citet{Poggianti_2009} find that galaxies residing in cluster environments might be more bulge-dominated. However, \citet{Allen_2016} find that only star-forming galaxies found in clusters are more likely to have higher S\'{e}rsic indices than their field counterparts but they report no difference among quiescent galaxies. 

In another study \citet{Bassett_2013} find no difference between field and cluster star-forming galaxies, but they reported that quiescent galaxies have a lower S\'{e}rsic index in a cluster environment. Galaxies with a high S\'{e}rsic index are also more abundant in denser cosmic neighbourhoods up to approximately $z \sim 1$ \citep{Postman_2005}. A study by \citet{Bruce_2014} delves deeper into the decomposition of light profiles, particularly focusing on bulge-to disc ratios, revealing an increasing dominance of bulges at around $z \sim 3$. There are indications in the current Universe that the formation of galactic bulges is particularly pronounced in regions of heightened density \citep{Lackner2012}. At intermediate redshifts $z \sim 0.4 - 0.8)$, \citet{Grossi_2017} observed a trend where galaxies with strong H$\alpha$ emission lines, which indicates young, star-forming galaxies, tend to possess more conspicuous bulges in denser environments. However, there remains a gap in our understanding concerning how the prevalence of bulges varies with the environment at these redshifts for samples selected via continuum methods.

\section{Discussion}\label{discussion}

\subsection{Comparison to other Works}

\subsubsection{Comparison to Carnall et al. (2023)}\label{comparingcarnall}

We perform a more detailed comparison of our treatment of objects in the CEERS field to the previous analysis done by \citet{carnall2023}. This analysis identified 15 massive galaxies as quiescent in this work. We note that 5 of the quenched objects identified do not appear in our final passive sample. 

Here we refer to objects using the ID numbers assigned in \citet{carnall2023} for ease of comparison. Galaxies 36262 and 80785 were flagged as having too low redshift to pass cuts imposed on the initial \texttt{EAZY} and the \texttt{BAGPIPES} samples respectively. The remaining three objects, 40015, 44362, and 28316, were cut due to high \texttt{BAGPIPES} sSFR values indicating they are likely still star-forming. 

Disagreement is perhaps not surprising, as significant changes and improvements have been made to the calibration and precision of the photometry used for SED fitting since this study, among the first conducted with JWST, was completed. was conducted. The reductions used in \citet{carnall2023}  used an adapted version of the CRDS version 0942, back when significant zero point offsets were present (e.g. \citep{adams2023discovery}). In our work, we utilise version 1084, which has implemented two incremental improvements to the NIRCam zero points since and should be accurate to the 5\% level.

\subsubsection{Comparison to Looser et al. (2024)}

The object found in \citet{looser2024recently} is debated as being the highest redshift confirmed passive object. We identify it in our catalogue with mostly similar fit parameters as derived by \citet{looser2024recently}. However, we reject it for our quenched sample after applying cuts to the \texttt{Le PHARE} data, as it has too high of a SFR to pass our quiescence criterion. Ultimately, we conclude that this disagreement is expected, as we use a different method of establishing quiescence, and the object may still have lingering blue stars, as it has only been passive for 10 - 100 Myr. \citet{looser2024recently} confirms quiescence through spectroscopy, observing that the spectrum of the galaxy appears flat in the H$\alpha$ and H$\beta$ regions. \citet{looser2024recently} suggests that the quenching observed may only be temporary, as it is located in a region sensitive to various feedback mechanisms.  

\subsubsection{Comparison to Kocevski et al. (2023)}

The intersection between objects that are passive and objects with an AGN is of great interest since AGNs are a source of quenching. We compare our matched catalogue findings with the table of passive objects containing AGN emission in \citet{kocevski2023ceers}. This work finds five objects detected within the CEERS field with \texttt{EAZY} fitted redshifts above $z=3$. We find all of them in the unfiltered \texttt{EAZY} data; however, we only identify two of these in our final passive sample. We find that 2 sources are lower redshift ($z < 2.5$, AEGIS-482 \& AEGIS-511) and that one target has too high of a sSFR (AEGIS-495). We find good agreement with the objects AEGIS 525 and 532 for both redshift and passive classification. 

The \citet{kocevski2023ceers} analysis uses NIRCam observations made using what turned out to be incorrect calibrations. Version 1.5.3 of the JWST Calibration Pipeline included the jwst$\_$nircam$\_$0214.imap NIRCAm reference files, the majority of which were created preflight. As detailed in \citet{adams2023discovery}, this initial set of calibrations had faults with the magnitude measurements in several bands, which likely impacted the quiescent selection process. These calibrations changed colours by up to $\sim$ 0.3 magnitudes, increasing the redder and decreasing the bluer bands. This shifting likely produced a greater Balmer break than would otherwise be present with current calibrations and thus produces a different selection in galaxy types. 

\subsection{Higher Redshift Galaxies Selected as Passive} \label{High z}

In our analysis, we have primarily focused on examining galaxies at $z<7$ though there are multiple galaxies detected towards higher redshifts. In total, there are 6 galaxies in the CEERS field at $z>6.5$, 0 in the JADES field and 0 in the NEP field. Above this redshift range, we find that all of our galaxies are very compact and red with approximately half of our candidates exhibiting SED shapes by eye that would be indicative of being classified as `little red dots', with flat/blue spectra in the rest-frame UV and a sharp red slope redwards of 3 microns. With this suspicion, we cross-match our sources to the LRD compilation of \citet{Kocevski2024}, finding that 3/6 galaxies at $z>6.5$ passing our selection criteria are contained within this compilation.

This is perhaps not too surprising, as the red colours of these sources lend themselves to being fit with dust and/or low star formation rates when fit with purely stellar templates. However, the debated contributions of AGN to the SED's of these sources could bias interpretations of the physical properties of these galaxies such as stellar mass and star formation rate. As such, we presently limit our discussion of the numbers of passive systems in this early epoch. Our 3 galaxies not initially identified as LRDs at $z>6.5$ are IDs 58926, 54880 and 79261 in Appendix A, all contained within the CEERS field.

\subsection{Potential for Future Work}\label{future}

There are several aspects in which cosmic variance in our sample could be minimised and the number densities and associated values could be improved. There are two major approaches to increasing the number of objects detected, namely including deeper fields or wider fields. Wider fields would also help to minimise the issue of cosmic variance, as this is a major factor when studying small fields. Since we have included JADES in our analysis, which is already deeper than we need for the detection of galaxies in our medium and large mass bins, we can conclude that future work efforts to better understand number densities will greatly benefit from including fields with larger areas. COSMOS-Web is a natural extension to this work, as the area, MIRI coverage and wealth of multiwavelength data, such as x-ray, radio and spectroscopy coverage, will greatly boost the understanding of passive systems.


One could also incorporate more detailed statistical and error considerations, specifically a more detailed discussion about redshift errors, and the range these cover compared to the fitted value itself. It is important to investigate how this factors into SED fit quality, or how it influences the number densities across different redshift bins.  

More and deeper radio data with e.g., the SKA or NGVLA would also help to resolve the nature of these quenched systems. 


In future work, we will investigate the nature of so-called naked cores and compact objects further. Work by many, including \citet{Carrasco2010} and \citet{costantin2022naked} suggest that most massive galaxies form from the inside outwards, by growing their extended star-forming disk around a central spheroid. As bulges tend to assemble mass at much quicker rates than disks, $\sim$ 0.7 versus $\sim$ 3.5 Gyr \citep[e.g.][]{margalef2016formation, costantin2022naked}, many compact objects we observe have the potential to be naked cores or the central bulges of galaxies that may develop a star-forming disk later. Follow-up analysis on these objects will be very important for improving understanding of galaxy evolution. 

\section{Conclusions} \label{conc}

In this work, we investigate the presence of massive quenched galaxies at redshifts higher than $z > 3$ in the CEERS, NEP and JADES fields, over a total area of 144.45 arcmin$^{2}$. We identify the features and number density of quiescent galaxies in the early universe as well as their stellar population, star formation history, and structure. Our major findings include:

\begin{itemize}

    \item We measure the evolution of quenched galaxy number densities over the redshift bins $3 \leq z < 4$, $4 \leq z < 5$ and $5 \leq z < 7$ within mass bins of $M \leq 10^{10.6} M_{\odot}$, $10^{9.5} M_{\odot} \leq M < 10^{10.6} M_{\odot}$), and $M \leq 10^{9.5} M_{\odot}$ which are listed in Table \ref{table:RESULTS} and Figure \ref{fig:evo1}. We find number densities of massive passive systems that are consistent with other observational studies \citep[e.g.][]{Carnall2020,Valentino_2023} and slightly higher number densities of moderately massive galaxies ($10^{9.5} M_{\odot} \leq M < 10^{10.6} M_{\odot}$). Numbers of passive galaxies are found to decrease by of order 1 dex between our $3<z<4$ bin and $5<z<7$ bins.
    
    \item We observe a significantly higher fraction of galaxies found to be quenched with our sSFR cut method than in work done with UVJ selection criteria. This is because UVJ sections identify galaxies which have been quenched for longer time periods \citep[see][]{trussler2024like}. We conclude that sSFR-based selection is required for higher redshift galaxy searches, as it is able to capture more recently quenched galaxies, where the UVJ based method requires both require long timescales post-quenching and, ideally, deep mid-infrared data.
    
    \item We find that our t-test results return no statistical difference between the sizes of the quiescent galaxy population and that of typical galaxies; however, a histogram of the distribution indicates that quenched galaxies are slightly more compact on average. We report a statistical difference between the S\'{e}rsic indices of the two populations (with passive galaxies having higher $n$), as represented in Figure \ref{fig:sersic_histogram}. Our observation of mostly compact bright objects with steep light profiles was reflected in the visual analysis and morphological classification, as reported in Table \ref{table:types} and Figure \ref{fig:mosaic}. We also confirm the presence of “red dots”, past and ongoing galaxy mergers, and pairs of galaxies in the sample. These present interesting opportunities for further study.
    
    \item Reviewing multi-wavelength data reveals a few sources have radio or x-ray detections, evidencing potential ongoing AGN activity in a small fraction of our sample. We find the quiescence of several quenched candidates to be reinforced by flat SEDs in the MIR data where it exists and remove three galaxies found to be dust-reddened instead of passive. We also find several objects with measured spectroscopy and suggest that further deeper coverage and red selections are needed to obtain statistical sample of passive systems.

    \item We find quenched galaxies using our definition of the specific star formation rate at all redshifts we probe. However, the highest redshift sources tend to exhibit LRD-like SED's making conclusions in this early time difficult until these sources are better understood.
\end{itemize}

\section*{Acknowledgements}

We acknowledge support from the ERC Advanced Investigator Grant EPOCHS (788113), as well as three studentships from the STFC. R.A.W., S.H.C., and R.A.J. acknowledge support from NASA JWST Interdisciplinary Scientist grants NAG5 12460, NNX14AN10G and 80NSSC18K0200 from GSFC. CNAW acknowledges funding from the JWST/NIRCam contract NASS-0215 to the University of Arizona. MAM acknowledges the support of a National Research Council of Canada Plaskett Fellowship, and the Australian Research Council Centre of Excellence for All Sky Astrophysics in 3 Dimensions (ASTRO 3D), through project number CE17010001. 

We thank the JADES, CEERS and PEARLS teams for their work in designing and preparing these public and GTO observations, and the STScI staff that carried them out. This work is based on observations made with the NASA/ESA \textit{Hubble Space Telescope} (HST) and NASA/ESA/CSA \textit{James Webb Space Telescope} (JWST) obtained from the \texttt{Mikulski Archive for Space Telescopes} (\texttt{MAST}) at the \textit{Space Telescope Science Institute} (STScI), which is operated by the Association of Universities for Research in Astronomy, Inc., under NASA contract NAS 5-03127 for JWST, and NAS 5–26555 for HST. Some of The data products presented herein were retrieved from the Dawn JWST Archive (DJA). DJA is an initiative of the Cosmic Dawn Center (DAWN), which is funded by the Danish National Research Foundation under grant DNRF140.

This work makes use of {\tt astropy} \citep{Astropy2013,Astropy2018,Astropy2022}, {\tt matplotlib} \citep{Hunter2007}, {\tt reproject}, {\tt DrizzlePac} \citep{Hoffmann2021}, {\tt SciPy} \citep{2020SciPy-NMeth} and {\tt photutils} \citep{larry_bradley_2022_6825092}.

\bibliographystyle{mnras}
\bibliography{Bibliography}

\appendix

\section{Full Table}\label{Appendix}

%
\clearpage


%

\begin{onecolumn}

\global\pdfpageattr\expandafter{\the\pdfpageattr/Rotate 90}

\begin{landscape}
\begin{longtable}{lllllllllllll}

\caption{The properties of the quiescent galaxies at $z < 3$ identified in this research. The properties of the galaxies include their coordinates, magnitude in the F444W band, photometric redshift found by \texttt{BAGPIPES}, total stellar mass, sSFR, time of formation and time of quenching found by \texttt{BAGPIPES}. The galaxies' radii and Sérsic Index determined by \texttt{GALFIT} are also included. Galaxies belonging to the robust passive sample have been marked with R after their ID.}
\label{tab:full_data} \\

\hline
\textbf{FIELD} & \textbf{ID} & \textbf{RA ($^\circ$)} & \textbf{DEC ($^\circ$)} & \textbf{F444W} & \textbf{$z_{\rm phot}$} & \textbf{$\log{M/M_\odot}$} & \textbf{log sSFR} & \textbf{\begin{tabular}[c]{@{}l@{}}$t_{form}$\\ (Gyr)\end{tabular}} & \textbf{\begin{tabular}[c]{@{}l@{}}$t_{quench}$\\ (Gyr)\end{tabular}} & \textbf{\begin{tabular}[c]{@{}l@{}}Radius\\ (kpc)\end{tabular}} & \textbf{\begin{tabular}[c]{@{}l@{}}Sérsic\\ index\end{tabular}} & \\
\hline
\endfirsthead

\multicolumn{13}{c}{Table \thetable{} -- Continued from previous page} \\
\hline
\textbf{FIELD} & \textbf{ID} & \textbf{RA ($^\circ$)} & \textbf{DEC ($^\circ$)} & \textbf{F444W} & \textbf{$z_{\rm phot}$} & \textbf{$\log{M/M_\odot}$} & \textbf{log sSFR} & \textbf{\begin{tabular}[c]{@{}l@{}}$t_{form}$\\ (Gyr)\end{tabular}} & \textbf{\begin{tabular}[c]{@{}l@{}}$t_{quench}$\\ (Gyr)\end{tabular}} & \textbf{\begin{tabular}[c]{@{}l@{}}Radius\\ (kpc)\end{tabular}} & \textbf{\begin{tabular}[c]{@{}l@{}}Sérsic\\ index\end{tabular}} & \\
\hline
\endhead

\hline
\multicolumn{13}{r}{Continued on next page} \\
\endfoot

\hline
\endlastfoot


CEERS & 1071  & 215.0031 & 53.0137  & $25.22_{-0.01}^{+0.01}$    & $3.55_{-0.27}^{+0.26}$ & $8.93_{-0.06}^{+0.05}$ & $-9.74_{-4.27}^{+1.85}$   & $1.63_{-0.13}^{+0.16}$ & $99_{-97.40}^{+0.00}$   & $1.83_{-0.03}^{+0.03}$ & $1.20_{-0.04}^{+0.04}$    \\
CEERS & 1229-R  & 214.9579 & 52.9803  & $22.92_{-0.00}^{+0.00}$    & $3.48_{-0.13}^{+0.18}$ & $10.39_{-0.04}^{+0.05}$ & $-61.44_{-37.71}^{+31.91}$   & $1.32_{-0.15}^{+0.13}$ & $1.35_{-0.15}^{+0.14}$   & $0.55_{-0.00}^{+0.00}$ & $1.44_{-0.02}^{+0.02}$    \\


CEERS & 2154-R  & 214.9818 & 52.9912  & $22.13_{-0.00}^{+0.00}$    & $3.47_{-0.15}^{+0.20}$ & $10.67_{-0.04}^{+0.04}$ & $-75.97_{-54.39}^{+47.19}$   & $1.20_{-0.17}^{+0.13}$ & $1.24_{-0.18}^{+0.14}$   & $0.71_{-0.00}^{+0.00}$ & $1.79_{-0.01}^{+0.01}$    \\
CEERS & 7088-R  & 215.0391 & 53.0028  & $23.37_{-0.01}^{+0.01}$    & $4.08_{-2.93}^{+0.22}$ & $10.45_{-0.69}^{+0.06}$ & $-118.40_{-...}^{+83.94}$    & $0.84_{-0.32}^{+1.39}$ & $0.87_{-0.34}^{+1.54}$   & $0.68_{-0.01}^{+0.01}$ & $1.07_{-0.07}^{+0.07}$    \\
CEERS & 11049 & 214.9022 & 52.9394  & $26.58_{-0.01}^{+0.01}$    & $9.95_{-0.44}^{+0.19}$ & $9.59_{-0.17}^{+0.14}$  & $-56.48_{-68.94}^{+42.58}$   & $0.29_{-0.06}^{+0.08}$ & $0.30_{-0.06}^{+0.08}$   & $0.22_{-0.03}^{+0.03}$ & $0.07_{-1.01}^{+1.01}$    \\
CEERS & 11746-R & 214.9049 & 52.9353  & $22.75_{-0.00}^{+0.00}$    & $3.34_{-0.11}^{+0.21}$ & $10.40_{-0.04}^{+0.04}$ & $-48.25_{-31.80}^{+24.63}$   & $1.43_{-0.13}^{+0.11}$ & $1.47_{-0.14}^{+0.12}$   & $0.58_{-0.00}^{+0.00}$ & $1.68_{-0.02}^{+0.02}$    \\
CEERS & 12017 & 214.8752 & 52.9135  & $23.65_{-0.00}^{+0.00}$    & $3.15_{-0.12}^{+0.13}$ & $9.87_{-0.06}^{+0.06}$  & $-19.29_{-16.57}^{+8.58}$    & $1.78_{-0.14}^{+0.12}$ & $1.82_{-0.13}^{+0.23}$   & $0.95_{-0.00}^{+0.00}$ & $1.55_{-0.02}^{+0.02}$    \\
CEERS & 12500-R & 214.9111 & 52.9331  & $21.15_{-0.00}^{+0.00}$    & $3.03_{-0.11}^{+0.09}$ & $10.86_{-0.05}^{+0.04}$ & $-23.65_{-14.79}^{+10.24}$   & $1.83_{-0.10}^{+0.12}$ & $1.87_{-0.11}^{+0.13}$   & $0.77_{-0.00}^{+0.00}$ & $1.70_{-0.01}^{+0.01}$    \\
CEERS & 12942 & 214.8583 & 52.8951  & $24.17_{-0.01}^{+0.01}$    & $4.19_{-3.20}^{+0.16}$ & $9.53_{-1.25}^{+0.05}$  & $-24.79_{-19.87}^{+16.67}$   & $1.23_{-0.06}^{+4.48}$ & $1.24_{-0.06}^{+97.76}$  & $1.46_{-0.03}^{+0.03}$ & $1.97_{-0.04}^{+0.04}$    \\
CEERS & 15660-R & 214.866  & 52.8841  & $22.26_{-0.00}^{+0.00}$    & $3.41_{-0.15}^{+0.31}$ & $10.54_{-0.05}^{+0.06}$ & $-36.19_{-35.87}^{+18.31}$   & $1.49_{-0.24}^{+0.15}$ & $1.52_{-0.24}^{+0.16}$   & $1.22_{-0.00}^{+0.00}$ & $1.44_{-0.01}^{+0.01}$    \\
CEERS & 15790-R & 214.8661 & 52.8843  & $21.83_{-0.00}^{+0.00}$    & $3.47_{-0.14}^{+0.19}$ & $10.80_{-0.04}^{+0.05}$ & $-63.21_{-40.97}^{+37.01}$   & $1.29_{-0.09}^{+0.10}$ & $1.33_{-0.11}^{+0.11}$   & $0.68_{-0.00}^{+0.00}$ & $1.74_{-0.02}^{+0.02}$    \\

CEERS & 17021  & 214.9440 & 52.9298  & $26.74_{-0.01}^{+0.01}$  & $3.55_{-0.27}^{+0.26}$ & $8.93_{-0.53}^{+0.07}$ & $-21.44_{-20.12}^{+10.54}$   & $1.63_{-0.13}^{+0.16}$ & $1.18_{-0.10}^{+4.57}$   & ... & ...    \\

CEERS & 17304  & 214.9310 & 52.9221  & $25.74_{-0.02}^{+0.02}$ & $3.24_{-0.13}^{+0.22}$ & $8.88_{-0.04}^{+0.04}$ & $-11.84_{-4.05}^{+2.69}$ & $1.78_{-0.12}^{+0.10}$ & $1.86_{-0.12}^{+97.13}$   & $0.66_{-0.02}^{+0.02}$ & $1.96_{-0.18}^{+0.18}$    \\

CEERS & 17349 & 214.8894 & 52.8922  & $25.77_{-0.02}^{+0.02}$    & $4.41_{-3.33}^{+0.18}$ & $9.12_{-0.06}^{+0.06}$  & $-15.60_{-8.74}^{+5.18}$     & $1.71_{-0.18}^{+0.11}$ & $1.76_{-0.19}^{+0.18}$   & $0.85_{-0.02}^{+0.02}$ & $1.63_{-0.13}^{+0.13}$    \\
CEERS & 17580-R & 214.8786 & 52.8882  & $24.97_{-0.03}^{+0.03}$    & $3.18_{-2.74}^{+0.17}$ & $9.24_{-0.98}^{+0.05}$  & $-55.67_{-71.17}^{+33.80}$   & $1.55_{-0.25}^{+1.11}$ & $1.59_{-0.27}^{+1.38}$   & $1.30_{-0.07}^{+0.07}$ & $3.52_{-0.21}^{+0.21}$    \\
CEERS & 17681-R & 214.8791 & 52.8881  & $23.27_{-0.00}^{+0.00}$    & $3.36_{-0.08}^{+0.13}$ & $10.19_{-0.04}^{+0.04}$ & $-33.80_{-23.40}^{+15.82}$   & $1.54_{-0.11}^{+0.11}$ & $1.57_{-0.12}^{+0.11}$   & $0.49_{-0.00}^{+0.00}$ & $1.77_{-0.06}^{+0.06}$    \\
CEERS & 18901-R & 214.7606 & 52.8453  & $20.92_{-0.00}^{+0.00}$    & $3.13_{-0.10}^{+0.10}$ & $10.95_{-0.04}^{+0.04}$ & $-63.16_{-47.81}^{+33.44}$   & $1.47_{-0.20}^{+0.14}$ & $1.51_{-0.21}^{+0.15}$   & $1.69_{-0.01}^{+0.01}$ & $3.77_{-0.03}^{+0.03}$    \\
CEERS & 21581-R & 214.7598 & 52.8334  & $25.59_{-0.01}^{+0.01}$    & $4.51_{-2.78}^{+0.20}$ & $9.70_{-0.46}^{+0.07}$  & $-99.94_{-174.51}^{+62.88}$  & $0.84_{-0.39}^{+0.29}$ & $0.87_{-0.41}^{+0.35}$   & $0.33_{-1.16}^{+1.16}$ & $0.05_{-0.89}^{+0.89}$    \\
CEERS & 22786-R & 214.7666 & 52.8315  & $25.41_{-0.01}^{+0.01}$    & $4.62_{-0.12}^{+0.14}$ & $9.63_{-0.06}^{+0.06}$  & $-42.70_{-42.51}^{+22.76}$   & $0.97_{-0.11}^{+0.11}$ & $0.99_{-0.12}^{+0.11}$   & $0.28_{-2.33}^{+2.33}$ & $0.05_{-1.54}^{+1.54}$    \\
CEERS & 23490-R & 214.811 & 52.8589  & $25.41_{-0.01}^{+0.01}$ & $3.02_{-2.02}^{+0.68}$ & $9.53_{-0.58}^{+0.09}$  & $-133.08_{-...}^{+97.15}$   & $1.03_{-0.50}^{+1.28}$ & $1.07_{-0.52}^{+1.31}$   & $1.12_{-0.02}^{+0.02}$ & $2.17_{-0.08}^{+0.08}$     \\

CEERS & 24636 & 214.8397 & 52.8723  & $25.61_{-0.02}^{+0.02}$ & $3.19_{-1.48}^{+0.11}$ & $8.81_{-0.08}^{+0.04}$  & $-21.25_{-16.32}^{+8.17}$ & $1.75_{-0.11}^{+0.18}$ & $1.80_{-0.11}^{+0.18}$  & $1.36_{-0.05}^{+0.05}$ & $1.69_{-0.09}^{+0.09}$    \\

CEERS & 24830 & 214.8397 & 52.8717  & $24.41_{-0.00}^{+0.00}$    & $4.00_{-0.47}^{+0.18}$ & $9.75_{-0.11}^{+0.06}$  & $-11.42_{-11.54}^{+3.53}$    & $1.39_{-0.12}^{+0.26}$ & $1.46_{-0.17}^{+97.54}$  & $0.60_{-0.01}^{+0.01}$ & $1.81_{-0.05}^{+0.05}$    \\
CEERS & 25071-R & 214.7672 & 52.8177  & $22.90_{-0.00}^{+0.00}$    & $3.71_{-0.32}^{+0.20}$ & $10.24_{-0.05}^{+0.05}$ & $-89.67_{-72.92}^{+53.51}$   & $1.05_{-0.21}^{+0.21}$ & $1.08_{-0.21}^{+0.22}$   & $0.82_{-0.00}^{+0.00}$ & $3.48_{-0.05}^{+0.05}$    \\
CEERS & 26564-R & 214.8539 & 52.8614  & $21.54_{-0.00}^{+0.00}$    & $3.48_{-0.25}^{+0.27}$ & $11.09_{-0.08}^{+0.06}$ & $-161.83_{-...}^{+102.67}$   & $0.77_{-0.40}^{+0.34}$ & $0.79_{-0.41}^{+0.36}$   & $1.37_{-0.00}^{+0.00}$ & $2.52_{-0.01}^{+0.01}$    \\
CEERS & 27157-R & 214.8506 & 52.8660   & $22.65_{-0.00}^{+0.00}$    & $3.72_{-2.58}^{+0.40}$ & $10.65_{-0.64}^{+0.11}$ & $-182.99_{-...}^{+126.44}$   & $0.73_{-0.32}^{+1.80}$ & $0.75_{-0.33}^{+1.83}$   & $0.59_{-0.00}^{+0.00}$ & $1.65_{-0.02}^{+0.02}$    \\
CEERS & 27958 & 214.8580  & 52.8763  & $25.22_{-0.03}^{+0.03}$    & $3.56_{-0.20}^{+0.18}$ & $8.90_{-0.04}^{+0.04}$  & $-20.11_{-8.60}^{+9.61}$     & $1.58_{-0.11}^{+0.11}$ & $1.61_{-0.12}^{+0.17}$   & $2.24_{-0.11}^{+0.11}$ & $2.03_{-0.10}^{+0.10}$    \\
CEERS & 28749-R & 214.8082 & 52.8322  & $24.28_{-0.00}^{+0.00}$    & $4.50_{-0.16}^{+0.14}$ & $10.25_{-0.05}^{+0.05}$ & $-63.66_{-72.34}^{+37.69}$   & $0.91_{-0.16}^{+0.13}$ & $0.93_{-0.16}^{+0.15}$   & $0.29_{-0.01}^{+0.01}$ & $0.25_{-0.15}^{+0.15}$    \\
CEERS & 30600-R & 214.7870  & 52.7708  & $22.82_{-0.00}^{+0.00}$    & $5.49_{-0.27}^{+0.18}$ & $10.64_{-0.08}^{+0.08}$ & $-45.60_{-47.44}^{+28.23}$   & $0.77_{-0.07}^{+0.07}$ & $0.79_{-0.08}^{+0.08}$   & $0.83_{-0.00}^{+0.00}$ & $1.33_{-0.01}^{+0.01}$    \\


CEERS & 32565-R & 214.7797 & 52.7776  & $22.46_{-0.00}^{+0.00}$    & $5.31_{-0.30}^{+0.24}$ & $10.72_{-0.06}^{+0.05}$ & $-25.04_{-18.07}^{+11.77}$   & $0.89_{-0.07}^{+0.08}$ & $0.92_{-0.08}^{+0.09}$   & $0.38_{-0.00}^{+0.00}$ & $1.75_{-0.01}^{+0.01}$    \\
CEERS & 36821-R & 214.7074 & 52.7526  & $22.61_{-0.00}^{+0.00}$    & $3.25_{-0.09}^{+0.10}$ & $10.43_{-0.06}^{+0.05}$ & $-25.35_{-21.91}^{+10.70}$   & $1.67_{-0.13}^{+0.12}$ & $1.71_{-0.14}^{+0.12}$   & $0.52_{-0.00}^{+0.00}$ & $1.64_{-0.02}^{+0.02}$    \\
CEERS & 37228 & 214.7189 & 52.7644  & $25.38_{-0.01}^{+0.01}$    & $5.64_{-0.50}^{+0.27}$ & $9.73_{-0.08}^{+0.09}$  & $-23.79_{-35.30}^{+15.10}$   & $0.82_{-0.13}^{+0.15}$ & $0.84_{-0.14}^{+98.16}$  & $0.47_{-0.01}^{+0.01}$ & $1.01_{-0.06}^{+0.06}$    \\

CEERS & 38757 & 214.9299 & 52.8622  & $23.49_{-0.01}^{+0.01}$    & $4.92_{-0.08}^{+0.08}$ & $9.95_{-0.05}^{+0.07}$  & $-12.89_{-12.88}^{+5.00}$ & $1.04_{-0.05}^{+0.06}$ & $1.07_{-0.06}^{+97.93}$  & $0.67_{-0.01}^{+0.01}$ & $3.89_{-0.09}^{+0.09}$    \\ 


CEERS & 43098 & 214.9295 & 52.8879  & $26.14_{-0.22}^{+0.22}$    & $8.99_{-0.23}^{+0.24}$ & $9.99_{-0.13}^{+0.15}$  & $-136.49_{-146.76}^{+82.65}$ & $0.22_{-0.07}^{+0.07}$ & $0.23_{-0.08}^{+0.07}$   & $0.25_{-0.03}^{+0.03}$ & $0.10_{-0.47}^{+0.47}$    \\

CEERS & 43971 & 214.9494 & 52.9079 & $24.36_{-0.01}^{+0.01}$    & $4.90_{-0.12}^{+0.15}$ & $9.54_{-0.04}^{+0.04}$  & $-14.45_{-7.85}^{+5.42}$ & $1.05_{-0.05}^{+0.04}$ & $1.08_{-0.05}^{+97.92}$   & $1.57_{-0.01}^{+0.01}$ & $1.0_{-0.02}^{+0.02}$    \\

CEERS & 45647 & 214.8922 & 52.8774  & $24.12_{-0.00}^{+0.00}$    & $7.18_{-0.15}^{+0.18}$ & $10.47_{-0.08}^{+0.09}$ & $-72.76_{-56.16}^{+46.03}$   & $0.46_{-0.06}^{+0.04}$ & $0.47_{-0.06}^{+0.05}$   & $0.27_{-0.01}^{+0.01}$ & $0.15_{-0.11}^{+0.11}$    \\
CEERS & 46407 & 214.9495 & 52.9139  & $21.51_{-0.00}^{+0.00}$    & $5.51_{-0.15}^{+0.13}$ & $11.24_{-0.11}^{+0.07}$ & $-22.74_{-36.46}^{+14.41}$   & $0.84_{-0.09}^{+0.08}$ & $0.87_{-0.10}^{+98.13}$  & $0.83_{-0.00}^{+0.00}$ & $2.52_{-0.02}^{+0.02}$    \\
CEERS & 46917 & 214.9659 & 52.9332  & $25.85_{-0.01}^{+0.01}$    & $6.01_{-0.20}^{+0.21}$ & $9.31_{-0.06}^{+0.06}$  & $-20.78_{-18.96}^{+10.79}$   & $0.75_{-0.06}^{+0.07}$ & $0.77_{-0.07}^{+0.16}$   & $0.62_{-0.01}^{+0.01}$ & $0.73_{-0.05}^{+0.05}$    \\
CEERS & 54880-R & 214.8494 & 52.8118  & $25.31_{-0.01}^{+0.01}$    & $6.70_{-0.19}^{+0.21}$ & $9.86_{-0.08}^{+0.10}$  & $-49.30_{-56.04}^{+29.75}$   & $0.55_{-0.08}^{+0.07}$ & $0.57_{-0.09}^{+0.07}$   & $0.18_{-0.00}^{+0.00}$ & $3.15_{-0.42}^{+0.42}$    \\

CEERS & 58926 & 215.1569 & 52.9708  & $26.03_{-0.02}^{+0.02}$    & $6.56_{-0.31}^{+0.20}$ & $9.16_{-0.07}^{+0.06}$  & $-10.16_{-10.34}^{+2.29}$   & $0.71_{-0.04}^{+0.06}$ & $0.78_{-0.10}^{+98.22}$   & $1.13_{-0.23}^{+0.23}$ & $0.05_{-0.05}^{+0.05}$    \\


CEERS & 68365 & 215.0128 & 52.8709  & $22.51_{-0.00}^{+0.00}$    & $3.08_{-0.52}^{+0.18}$ & $10.48_{-0.16}^{+0.09}$ & $-11.48_{-24.70}^{+3.58}$    & $1.92_{-0.19}^{+0.23}$ & $2.58_{-0.80}^{+96.42}$  & $0.98_{-0.00}^{+0.00}$ & $1.14_{-0.01}^{+0.01}$    \\
CEERS & 70854-R & 215.0264 & 52.8938  & $22.78_{-0.00}^{+0.00}$    & $4.25_{-0.13}^{+0.13}$ & $10.67_{-0.05}^{+0.04}$ & $-109.43_{-90.22}^{+68.51}$  & $0.84_{-0.19}^{+0.15}$ & $0.85_{-0.20}^{+0.16}$   & $0.44_{-0.00}^{+0.00}$ & $1.33_{-0.02}^{+0.02}$    \\
CEERS & 74351 & 215.0099 & 52.9107  & $24.73_{-0.01}^{+0.01}$    & $4.96_{-0.15}^{+0.14}$ & $9.52_{-0.05}^{+0.05}$  & $-13.31_{-11.41}^{+5.24}$    & $1.03_{-0.05}^{+0.06}$ & $1.06_{-0.06}^{+97.94}$  & $0.94_{-0.01}^{+0.01}$ & $1.37_{-0.03}^{+0.03}$    \\
CEERS & 75219 & 214.9813 & 52.8826  & $24.13_{-0.01}^{+0.01}$    & $4.81_{-0.15}^{+0.25}$ & $10.03_{-0.07}^{+0.07}$ & $-35.68_{-36.98}^{+18.86}$   & $0.94_{-0.12}^{+0.10}$ & $0.96_{-0.11}^{+0.11}$   & $0.91_{-0.01}^{+0.01}$ & $2.67_{-0.06}^{+0.06}$    \\

CEERS & 76188 & 215.006  & 52.9053  & $24.40_{-0.01}^{+0.01}$    & $5.56_{-0.17}^{+0.17}$ & $9.79_{-0.05}^{+0.05}$  & $-24.10_{-18.70}^{+10.93}$   & $0.83_{-0.04}^{+0.05}$ & $0.85_{-0.05}^{+0.06}$   & $0.74_{-0.01}^{+0.01}$ & $2.37_{-0.08}^{+0.08}$    \\
CEERS & 78030 & 214.897  & 52.7922  & $22.88_{-0.00}^{+0.00}$    & $3.29_{-0.11}^{+0.18}$ & $10.29_{-0.06}^{+0.07}$ & $-11.83_{-11.11}^{+3.86}$    & $1.74_{-0.13}^{+0.13}$ & $1.81_{-0.15}^{+97.19}$  & $0.86_{-0.00}^{+0.00}$ & $1.20_{-0.01}^{+0.01}$    \\
CEERS & 79261-R & 214.9118 & 52.8091  & $24.92_{-0.00}^{+0.00}$    & $6.73_{-0.17}^{+0.16}$ & $10.02_{-0.07}^{+0.09}$ & $-72.51_{-60.16}^{+41.01}$   & $0.49_{-0.05}^{+0.06}$ & $0.50_{-0.06}^{+0.07}$   & $0.23_{-0.00}^{+0.00}$ & $0.83_{-0.08}^{+0.08}$    \\


CEERS & 82452-R & 214.8949 & 52.8172  & $22.92_{-0.00}^{+0.00}$    & $3.11_{-1.72}^{+0.34}$ & $10.46_{-0.32}^{+0.07}$ & $-181.83_{-nan}^{+123.04}$   & $0.83_{-0.42}^{+0.62}$ & $0.84_{-0.43}^{+0.68}$   & $0.80_{-0.00}^{+0.00}$ & $1.31_{-0.01}^{+0.01}$    \\
CEERS & 83148 & 214.9191 & 52.8402  & $22.12_{-0.01}^{+0.01}$    & $3.31_{-0.19}^{+0.69}$ & $9.88_{-0.09}^{+0.13}$  & $-18.02_{-19.25}^{+7.50}$    & $1.69_{-0.38}^{+0.16}$ & $1.74_{-0.41}^{+0.25}$   & $4.15_{-0.06}^{+0.06}$ & $1.93_{-0.03}^{+0.03}$    \\


CEERS & 90679 & 214.8496 & 52.7822  & $24.17_{-0.00}^{+0.00}$    & $3.21_{-0.12}^{+0.20}$ & $9.56_{-0.05}^{+0.05}$  & $-13.11_{-9.09}^{+4.49}$     & $1.80_{-0.14}^{+0.11}$ & $1.86_{-0.14}^{+97.14}$  & $0.71_{-0.00}^{+0.00}$ & $1.04_{-0.02}^{+0.02}$    \\

CEERS & 93088 & 214.8451 & 52.7918 & $25.15_{-0.01}^{+0.01}$   & $4.35_{-0.22}^{+0.20}$ & $9.18_{-0.06}^{+0.07}$  & $-12.06_{-8.57}^{+4.16}$     & $1.24_{-0.12}^{+0.12}$ & $1.28_{-0.13}^{+97.72}$  & $1.68_{-0.04}^{+0.04}$ & $1.43_{-0.05}^{+0.05}$    \\

CEERS & 93773 & 214.76538 & 52.7430  & $23.43_{-0.01}^{+0.01}$    & $3.01_{-0.14}^{+0.11}$ & $9.41_{-0.05}^{+0.05}$ & $-11.10_{-6.31}^{+3.03}$  & $1.96_{-0.09}^{+0.13}$ & $2.07_{-0.15}^{+96.93}$  & $2.84_{-0.03}^{+0.03}$ & $2.08_{-0.03}^{+0.03}$    \\

CEERS & 93869 & 214.7738 & 52.7400    & $22.73_{-0.00}^{+0.00}$    & $4.03_{-0.40}^{+0.23}$ & $10.09_{-0.11}^{+0.07}$ & $-16.10_{-27.89}^{+7.80}$    & $1.34_{-0.21}^{+0.25}$ & $1.40_{-0.24}^{+97.60}$  & $2.79_{-0.02}^{+0.02}$ & $1.81_{-0.01}^{+0.01}$    \\
NEP   & 1331-R  & 260.7022 & 65.7911  & $25.35_{-0.00}^{+0.00}$    & $4.51_{-0.20}^{+0.19}$ & $9.86_{-0.07}^{+0.06}$  & $-62.44_{-98.87}^{+38.33}$   & $0.90_{-0.28}^{+0.18}$ & $0.93_{-0.29}^{+0.19}$   & $0.26_{-0.02}^{+0.02}$ & $0.57_{-0.30}^{+0.30}$    \\


NEP   & 13593 & 260.7311 & 65.7417  & $23.62_{-0.00}^{+0.00}$    & $3.08_{-0.13}^{+0.12}$ & $9.56_{-0.06}^{+0.05}$  & $-12.49_{-8.31}^{+4.25}$     & $1.90_{-0.11}^{+0.13}$ & $1.96_{-0.14}^{+97.04}$  & $1.65_{-0.01}^{+0.01}$ & $2.01_{-0.02}^{+0.02}$    \\
NEP   & 27737-R & 260.8609 & 65.8295  & $25.61_{-0.00}^{+0.00}$    & $5.14_{-0.20}^{+0.25}$ & $9.51_{-0.06}^{+0.06}$  & $-30.50_{-25.20}^{+16.02}$   & $0.90_{-0.08}^{+0.07}$ & $0.92_{-0.09}^{+0.08}$   & $0.28_{-0.01}^{+0.01}$ & $0.06_{-0.63}^{+0.63}$    \\



NEP   & 40332-R & 260.7017 & 65.8669  & $24.04_{-0.01}^{+0.01}$    & $3.68_{-0.22}^{+0.24}$ & $10.07_{-0.05}^{+0.05}$ & $-87.12_{-114.11}^{+49.29}$  & $1.01_{-0.31}^{+0.22}$ & $1.04_{-0.31}^{+0.24}$   & $0.79_{-0.01}^{+0.01}$ & $2.87_{-0.08}^{+0.08}$    \\


NEP   & 40872 & 260.6987 & 65.8681  & $26.52_{-0.02}^{+0.02}$    & $3.61_{-0.17}^{+0.21}$ & $8.69_{-0.05}^{+0.05}$  & $-19.04_{-10.97}^{+8.23}$    & $1.52_{-0.13}^{+0.10}$ & $1.56_{-0.15}^{+0.14}$   & $1.01_{-0.03}^{+0.03}$ & $0.75_{-0.07}^{+0.07}$    \\

NEP   & 43043 & 260.7251 & 65.9197  & $23.59_{-0.00}^{+0.00}$    & $3.30_{-0.15}^{+0.26}$ & $9.75_{-0.05}^{+0.05}$  & $-10.57_{-7.08}^{+2.67}$    & $1.76_{-0.15}^{+0.12}$ & $1.91_{-0.22}^{+97.09}$   & $0.85_{-0.01}^{+0.01}$ & $2.79_{-0.05}^{+0.05}$    \\ 

NEP   & 44999-R & 260.6879 & 65.8843  & $23.33_{-0.01}^{+0.01}$    & $3.32_{-0.10}^{+0.18}$ & $10.04_{-0.04}^{+0.05}$ & $-85.83_{-67.05}^{+49.66}$   & $1.18_{-0.17}^{+0.16}$ & $1.22_{-0.16}^{+0.18}$   & $0.99_{-0.02}^{+0.02}$ & $4.72_{-0.10}^{+0.10}$    \\

NEP   & 45024 & 260.6876 & 65.8845  & $21.83_{-0.00}^{+0.00}$    & $3.16_{-0.12}^{+0.09}$ & $10.72_{-0.06}^{+0.06}$  & $-17.17_{-13.54}^{+7.75}$   & $1.80_{-0.11}^{+0.16}$ & $1.84_{-0.11}^{+97.16}$  & $0.64_{-0.00}^{+0.00}$ & $4.90_{-0.09}^{+0.09}$    \\

NEP   & 45091 & 260.6416 & 65.8224  & $23.79_{-0.01}^{+0.01}$    & $4.47_{-0.20}^{+0.14}$ & $9.56_{-0.06}^{+0.07}$  & $-19.35_{-18.98}^{+10.47}$   & $1.14_{-0.09}^{+0.12}$ & $1.17_{-0.09}^{+97.83}$  & $2.43_{-0.03}^{+0.03}$ & $1.60_{-0.03}^{+0.03}$    \\
NEP   & 45354-R & 260.6872 & 65.8844  & $23.27_{-0.01}^{+0.01}$    & $3.13_{-0.21}^{+0.10}$ & $9.83_{-0.08}^{+0.04}$  & $-95.57_{-86.97}^{+56.42}$   & $1.25_{-0.22}^{+0.31}$ & $1.30_{-0.26}^{+0.29}$   & $2.70_{-0.05}^{+0.05}$ & $3.62_{-0.06}^{+0.06}$    \\

NEP   & 45841 & 260.7078 & 65.9143  & $23.81_{-0.00}^{+0.00}$    & $4.17_{-0.10}^{+0.10}$ & $9.8354_{-0.04}^{+0.04}$  & $-15.17_{-12.31}^{+5.97}$   & $1.29_{-0.06}^{+0.05}$ & $1.32_{-0.06}^{+97.68}$   & $0.90_{-0.01}^{+0.01}$ & $1.62_{-0.03}^{+0.03}$    \\

NEP   & 46844-R & 260.6631 & 65.8614  & $21.74_{-0.00}^{+0.00}$    & $3.96_{-0.34}^{+0.18}$ & $10.79_{-0.07}^{+0.08}$ & $-55.76_{-59.96}^{+33.04}$   & $1.11_{-0.23}^{+0.22}$ & $1.14_{-0.23}^{+0.22}$   & $1.32_{-0.00}^{+0.00}$ & $2.04_{-0.02}^{+0.02}$    \\


NEP   & 52689-R & 260.6702 & 65.9063  & $25.08_{-0.02}^{+0.02}$    & $3.82_{-0.51}^{+0.25}$ & $9.31_{-0.08}^{+0.06}$  & $-42.00_{-58.62}^{+22.90}$   & $1.25_{-0.18}^{+0.26}$ & $1.28_{-0.18}^{+0.27}$   & $1.30_{-0.03}^{+0.03}$ & $1.32_{-0.05}^{+0.05}$    \\


NEP   & 54448 & 260.6479 & 65.8871  & $22.52_{-0.00}^{+0.00}$    & $3.25_{-0.10}^{+0.11}$ & $10.40_{-0.06}^{+0.06}$ & $-12.09_{-9.75}^{+3.72}$     & $1.78_{-0.10}^{+0.10}$ & $1.84_{-0.12}^{+97.16}$  & $1.07_{-0.00}^{+0.00}$ & $1.02_{-0.01}^{+0.01}$    \\


NEP   & 55349-R & 260.6337 & 65.862   & $23.90_{-0.01}^{+0.01}$    & $4.34_{-0.21}^{+0.17}$ & $10.08_{-0.05}^{+0.05}$ & $-99.62_{-111.74}^{+60.25}$  & $0.82_{-0.22}^{+0.18}$ & $0.84_{-0.23}^{+0.20}$   & $0.94_{-0.01}^{+0.01}$ & $1.32_{-0.05}^{+0.05}$    \\
NEP   & 55884-R & 260.6142 & 65.8562  & $21.94_{-0.00}^{+0.00}$    & $3.40_{-0.11}^{+0.18}$ & $10.70_{-0.03}^{+0.05}$ & $-70.30_{-56.59}^{+42.27}$   & $1.27_{-0.18}^{+0.15}$ & $1.30_{-0.19}^{+0.16}$   & $0.78_{-0.00}^{+0.00}$ & $2.36_{-0.01}^{+0.01}$    \\
NEP   & 58270 & 260.4833 & 65.8381  & $24.01_{-0.01}^{+0.01}$    & $4.39_{-0.22}^{+0.17}$ & $9.72_{-0.07}^{+0.07}$  & $-19.33_{-24.58}^{+10.00}$   & $1.17_{-0.14}^{+0.14}$ & $1.21_{-0.15}^{+97.79}$  & $1.74_{-0.02}^{+0.02}$ & $1.11_{-0.02}^{+0.02}$    \\

NEP   & 62768 & 260.6597 & 65.8090  & $26.97_{-0.02}^{+0.02}$    & $5.11_{-0.20}^{+0.20}$ & $8.85_{-0.04}^{+0.06}$  & $-9.58_{-4.76}^{+1.70}$   & $1.00_{-0.05}^{+0.06}$ & $99.0_{-97.99}^{+0.00}$  & $0.40_{-0.13}^{+0.13}$ & $0.06_{-0.83}^{+0.83}$    \\

NEP   & 65420-R & 260.5629 & 65.8146  & $23.88_{-0.00}^{+0.00}$    & $4.07_{-0.15}^{+0.12}$ & $9.95_{-0.05}^{+0.05}$  & $-29.10_{-23.67}^{+13.39}$   & $1.24_{-0.08}^{+0.10}$ & $1.27_{-0.09}^{+0.10}$   & $0.62_{-0.00}^{+0.00}$ & $1.84_{-0.03}^{+0.03}$    \\
NEP   & 68590-R & 260.5059 & 65.815   & $23.39_{-0.00}^{+0.00}$    & $3.34_{-0.09}^{+0.11}$ & $10.25_{-0.03}^{+0.05}$ & $-80.27_{-71.13}^{+42.20}$   & $1.26_{-0.26}^{+0.16}$ & $1.30_{-0.27}^{+0.16}$   & $0.49_{-0.00}^{+0.00}$ & $1.09_{-0.02}^{+0.02}$    \\

NEP   & 69691 & 260.6075 & 65.8007  & $24.09_{-0.00}^{+0.00}$    & $3.61_{-0.25}^{+0.22}$ & $9.63_{-0.05}^{+0.04}$ & $-11.53_{-6.84}^{+3.46}$   & $1.58_{-0.10}^{+0.14}$ & $1.73_{-0.21}^{+97.27}$   & $0.71_{-0.00}^{+0.00}$ & $1.37_{-0.03}^{+0.03}$    \\

NEP   & 70650 & 260.5341 & 65.8062  & $22.43_{-0.01}^{+0.01}$    & $3.11_{-0.10}^{+0.09}$ & $10.11_{-0.06}^{+0.05}$ & $-19.25_{-14.62}^{+7.60}$    & $1.82_{-0.12}^{+0.12}$ & $1.86_{-0.13}^{+0.13}$   & $2.69_{-0.06}^{+0.06}$ & $4.76_{-0.09}^{+0.09}$    \\
NEP   & 71472-R & 260.5342 & 65.8058  & $24.43_{-0.02}^{+0.02}$    & $3.24_{-0.13}^{+0.15}$ & $9.22_{-0.04}^{+0.04}$  & $-23.74_{-9.21}^{+7.19}$     & $1.72_{-0.12}^{+0.07}$ & $1.75_{-0.13}^{+0.07}$   & $1.95_{-0.06}^{+0.06}$ & $2.77_{-0.08}^{+0.08}$    \\

NEP   & 71500 & 260.5333 & 65.8048  & $23.48_{-0.00}^{+0.00}$    & $3.23_{-0.23}^{+0.59}$ & $9.63_{-0.10}^{+0.11}$  & $-14.00_{-13.75}^{+5.35}$    & $1.78_{-0.36}^{+0.20}$ & $1.87_{-0.41}^{+97.13}$  & $2.52_{-0.01}^{+0.01}$ & $1.40_{-0.01}^{+0.01}$    \\
NEP   & 72863 & 260.6481 & 65.7817  & $23.72_{-0.01}^{+0.01}$    & $4.37_{-0.13}^{+0.13}$ & $9.75_{-0.06}^{+0.05}$  & $-33.43_{-29.90}^{+18.43}$   & $1.11_{-0.11}^{+0.08}$ & $1.14_{-0.11}^{+0.09}$   & $2.09_{-0.03}^{+0.03}$ & $1.50_{-0.02}^{+0.02}$    \\

NEP   & 74274 & 260.4795 & 65.8084  & $23.90_{-0.02}^{+0.02}$    & $4.23_{-0.18}^{+0.16}$ & $9.45_{-0.06}^{+0.06}$  & $-13.53_{-16.13}^{+5.61}$   & $1.27_{-0.10}^{+0.10}$ & $1.31_{-0.12}^{+97.69}$   & $10.25_{-0.35}^{+0.35}$ & $2.91_{-0.09}^{+0.09}$  \\

NEP   & 74400 & 260.5185 & 65.804   & $23.89_{-0.00}^{+0.00}$    & $3.32_{-0.14}^{+0.41}$ & $9.72_{-0.05}^{+0.06}$  & $-21.32_{-12.98}^{+9.31}$    & $1.64_{-0.19}^{+0.11}$ & $1.70_{-0.21}^{+0.12}$   & $0.81_{-0.00}^{+0.00}$ & $1.51_{-0.02}^{+0.02}$    \\
NEP   & 74840-R & 260.6018 & 65.7951  & $25.16_{-0.00}^{+0.00}$    & $4.65_{-0.12}^{+0.18}$ & $9.68_{-0.05}^{+0.07}$  & $-37.88_{-33.48}^{+19.28}$   & $0.99_{-0.09}^{+0.09}$ & $1.02_{-0.10}^{+0.09}$   & $0.25_{-0.01}^{+0.01}$ & $1.05_{-0.17}^{+0.17}$    \\
JADES & 11241-R & 53.1969  & -27.7605 & $22.63_{-0.00}^{+0.00}$    & $3.60_{-0.19}^{+0.25}$ & $10.62_{-0.05}^{+0.05}$ & $-109.10_{-89.70}^{+62.66}$  & $0.97_{-0.22}^{+0.18}$ & $1.01_{-0.24}^{+0.20}$   & $0.60_{-0.00}^{+0.00}$ & $1.07_{-0.01}^{+0.01}$    \\
JADES & 12118 & 53.1966  & -27.7571 & $23.69_{-0.00}^{+0.00}$    & $4.10_{-0.30}^{+0.18}$ & $10.04_{-0.07}^{+0.07}$ & $-23.59_{-52.51}^{+12.44}$   & $1.24_{-0.26}^{+0.23}$ & $1.27_{-0.28}^{+0.31}$   & $1.21_{-0.01}^{+0.01}$ & $1.56_{-0.01}^{+0.01}$    \\
JADES & 12135-R & 53.187   & -27.7756 & $27.32_{-0.02}^{+0.02}$    & $4.48_{-0.18}^{+0.11}$ & $8.82_{-0.07}^{+0.05}$  & $-47.86_{-46.78}^{+22.03}$   & $0.98_{-0.08}^{+0.15}$ & $1.00_{-0.10}^{+0.16}$   & $0.24_{-0.01}^{+0.01}$ & $2.55_{-0.72}^{+0.72}$    \\
JADES & 16097 & 53.187   & -27.7523 & $25.30_{-0.01}^{+0.01}$    & $3.44_{-0.16}^{+0.26}$ & $9.04_{-0.05}^{+0.06}$  & $-16.48_{-9.04}^{+6.29}$     & $1.65_{-0.13}^{+0.09}$ & $1.69_{-0.14}^{+0.22}$   & $1.44_{-0.03}^{+0.03}$ & $1.11_{-0.05}^{+0.05}$    \\


JADES & 18418-R & 53.1812  & -27.7565 & $22.74_{-0.00}^{+0.00}$    & $3.47_{-0.14}^{+0.27}$ & $10.54_{-0.05}^{+0.05}$ & $-126.55_{-99.99}^{+74.66}$  & $0.94_{-0.24}^{+0.29}$ & $0.97_{-0.25}^{+0.29}$   & $0.60_{-0.00}^{+0.00}$ & $1.90_{-0.02}^{+0.02}$    \\

JADES & 20369 & 53.1518  & -27.8002 & $28.97_{-0.03}^{+0.03}$    & $3.12_{-2.72}^{+0.31}$ & $7.61_{-0.91}^{+0.08}$ & $-13.39_{-48.98}^{+5.29}$  & $1.86_{-0.18}^{+2.91}$ & $5.24_{-3.47}^{+93.76}$   & $0.63_{-0.05}^{+0.05}$ & $0.67_{-0.38}^{+0.38}$    \\

JADES & 25772 & 53.1666  & -27.7534 & $25.85_{-0.01}^{+0.01}$    & $4.58_{-0.12}^{+0.18}$ & $8.66_{-0.05}^{+0.04}$ & $-15.06_{-7.96}^{+6.01}$  & $1.13_{-0.05}^{+0.06}$ & $1.18_{-0.06}^{+97.82}$   & $2.48_{-0.05}^{+0.05}$ & $1.15_{-0.03}^{+0.03}$    \\

JADES & 29438 & 53.1191  & -27.814  & $23.42_{-0.00}^{+0.00}$    & $3.94_{-0.26}^{+0.21}$ & $9.88_{-0.08}^{+0.07}$  & $-21.97_{-21.59}^{+11.02}$   & $1.34_{-0.15}^{+0.16}$ & $1.38_{-0.15}^{+0.24}$   & $2.84_{-0.01}^{+0.01}$ & $1.39_{-0.01}^{+0.01}$    \\
JADES & 30450-R & 53.1305  & -27.7912 & $23.71_{-0.00}^{+0.00}$    & $3.61_{-0.18}^{+0.17}$ & $10.05_{-0.06}^{+0.05}$ & $-26.10_{-18.27}^{+10.98}$   & $1.46_{-0.09}^{+0.12}$ & $1.50_{-0.10}^{+0.13}$   & $0.64_{-0.01}^{+0.01}$ & $2.32_{-0.03}^{+0.03}$    \\
JADES & 34464 & 53.1388  & -27.757  & $24.87_{-0.00}^{+0.00}$    & $3.17_{-0.14}^{+0.14}$ & $9.10_{-0.05}^{+0.05}$  & $-11.68_{-8.79}^{+3.53}$     & $1.84_{-0.13}^{+0.13}$ & $1.92_{-0.15}^{+97.08}$  & $1.65_{-0.01}^{+0.01}$ & $1.31_{-0.02}^{+0.02}$    \\



\end{longtable}
\end{landscape}

\end{onecolumn}

\end{document}